\documentclass[a4paper,fleqn]{cas-sc}

\usepackage[numbers]{natbib}
\usepackage{amsmath}
\usepackage{hyperref}
\usepackage{mathtools}
\usepackage{setspace}
\usepackage{booktabs}
\usepackage[ruled]{algorithm2e}[1]
\usepackage{graphicx}
\usepackage{longtable}
\usepackage{amssymb}
\usepackage{float}

\def\tsc#1{\csdef{#1}{\textsc{\lowercase{#1}}\xspace}}
\tsc{WGM}
\tsc{QE}
\tsc{EP}
\tsc{PMS}
\tsc{BEC}
\tsc{DE}

\begin{document}
\let\WriteBookmarks\relax
\def\floatpagepagefraction{1}
\def\textpagefraction{.001}

\shortauthors{Chen et~al.}

\shorttitle{ }

\title [mode = title]{Timestamp calibration for time-series single cell RNA-seq expression data}


%
\author[1]{Xiran Chen}


\credit{Methodology, Investigation, Software, Visualization, Writing – original draft}

\affiliation[1]{organization={College of mathematics and statistic, Chongqing Jiaotong University},
    city={Chongqing},
    country={China}}

\author[2]{Sha Lin}

\affiliation[2]{organization={Key Laboratory of Birth Defects and Related Diseases of Women and Children of MOE, Department of Pediatrics, West China Second University Hospital, Sichuan University},
    city={Chengdu},
    country={China}}
\credit{Validation, Writing – original draft}
\author[1]{Xiaofeng Chen}

\credit{Methodology, Supervision}

\author%
[1]
{Weikai Li}
\cormark[1]
\ead{leeweikai@outlook.com}

\credit{Methodology, Supervision, Writing – review and editing}

\author%
[2]
{Yifei Li}
\cormark[1]
\ead{liyfwcsh@scu.edu.cn}

\credit{Supervision, Writing – review and editing}

\cortext[cor1]{Corresponding author}

\begin{abstract}
Timestamp automatic annotation (TAA) is a crucial procedure for analyzing time-series ScRNA-seq data, as they unveil dynamic biological developments and cell regeneration process. However, current TAA methods heavily rely on manual timestamps, often overlooking their reliability. This oversight can significantly degrade the performance of timestamp automatic annotation due to noisy timestamps. Nevertheless, the current approach for addressing this issue tends to select less critical cleaned samples for timestamp calibration. To tackle this challenge, we have developed a novel timestamp calibration model called ScPace for handling noisy labeled time-series ScRNA-seq data. This approach incorporates a latent variable indicator within a base classifier instead of probability sampling to detect noisy samples effectively. To validate our proposed method, we conducted experiments on both simulated and real time-series ScRNA-seq datasets. Cross-validation experiments with different artificial mislabeling rates demonstrate that ScPace outperforms previous approaches. Furthermore, after calibrating the timestamps of the original time-series ScRNA-seq data using our method, we performed supervised pseudotime analysis, revealing that ScPace enhances its performance significantly. These findings suggest that ScPace is an effective tool for timestamp calibration by enabling reclassification and deletion of detected noisy labeled samples while maintaining robustness across diverse ranges of time-series ScRNA-seq datasets. The source code is available at https://github.com/OPUS-Lightphenexx/ScPace.
\end{abstract}




\maketitle

\section{Introduction}\label{intro}


Time-series ScRNA-seq data differ from traditional ScRNA-seq data, as cells are collected at distinct time points, it could provide insights into cellular cycles\cite{1}, developmental process\cite{2} and disease development\cite{disease}. Over the past decades, numerous cell type annotation methods have been developed\cite{cellanno}, ranging from unsupervised clustering\cite{nature_cluster} methods to supervised approaches\cite{scbert}. Specifically, timestamp automatic annotation has grown rapidly attention and numerous methods have been developed for this task\cite{devcellpy}. Particularly, this area of research is important in revealing how cardiomyocyte regenerate, which has been regarded as a promising way to affect the pathological process even in clinical practice.

The timestamps in time-series ScRNA-seq data are derived from the timing of sample collection. However, time-series ScRNA-seq datasets often exhibit mislabeled cells with noisy timestamps due to various factors, resulting in datasets that are compromised by labeling inaccuracies. Several reasons contribute to the mislabeling in time-series datasets. For instance, when different heterogeneous cell types are present at a single time point, misclassification may occur as cells are erroneously assigned to adjacent time points. Additionally, extreme technical dropouts in certain cells\cite{gangli}, and different maturation state at one timestamp can cause cell states to overlap with other distinct timestamps\cite{gse90047}, further complicating the ordinal relationships among timestamps. These factors could contribute in degrading the performance of timestamp automatic annotation.

To tackle this problem, the computational approach employed by ScReclassify\cite{screclassify} for the reclassification of mislabeled cell types utilizes a semi-supervised Adasampling technique, it could also be adapted for the calibration of timestamps. However, the existing mechanism for selecting potential mislabeled cells leverage probabilistic sampling, which may lead to selecting unimportant samples during the sample selection process. Moreover, ScReclassify does not adequately account for the intricate relationships inherent in time-series ScRNA-seq data. Specifically, in complex nonlinear time-series ScRNA-seq contexts, ScReclassify relies on principal component analysis (PCA) conducted on the initial dataset in all conditions, which may lead to information leakage, particularly in sparse ScRNA-seq datasets\cite{methods}. 

Therefore, we hypothesize that conventional timestamp automatic annotation, which rely heavily on high-quality timestamp labels to yield satisfactory performance, may experience decreased performance when applied to datasets with noisy timestamp labels, and current approach aimed at addressing this issue rely on probability sampling, which may result in the selection of unimportant samples. To mitigate these challenges, we developed a novel timestamp calibration method for time-series ScRNA-seq data, referred to as ScPace. In contrast to previous method that correct cell type labels using semi-supervised Adasampling techniques, our approach introduces a latent variable indicator for the selection of potentially mislabeled samples, enabling their deletion or reclassification according to specific requirements. This method is informed by the principles of self-paced learning\cite{spl}, a machine learning training paradigm that emulates the cognitive processes involved in human learning by progressively transitioning from easier to more difficult samples, which has being applied to various fields like multi-modal image classification\cite{spl_cla}, domain adaption\cite{spl_domain}, imbalance sampling\cite{spl_imbalance} and functional brain network scrubbing\cite{fbn}. Specially, previous research on imbalance sampling utilize self paced learning has achieved better performance than random imbalance sampling methods, which emphasizes the retention of high-quality samples while discarding bad-quality samples to construct a balanced samples.

To the best of our knowledge, ScPace is the first method specifically designed to focus on the calibration of timestamps within time-series ScRNA-seq datasets and assessing its enhancement of timestamp automatic annotation and supervised pseudotime analysis. In summary, the contributions of our proposed method can be listed as follows:
\begin{itemize}
    \item In contrast to prior method that address the calibration of cell type labels through semi-supervised Adasampling techniques which may overlook critical samples, we introduce a latent variable indicator designed for the selection of potentially mislabeled samples. Our proposed method facilitates the deletion or reclassification of such samples based on specific requirements. 
    \item Besides timestamp automatic annotation, Supervised pseudotime analysis utilizing timestamp labels is significantly dependent on the quality of the timestamps. Consequently, we further enhance performance following the calibration of the timestamps through our proposed method on the original time-series ScRNA-seq data. This correction contributes positively to subsequent downstream pseudotime analyses.
    \item  Our method demonstrates enhanced robustness in the context of imbalanced and high-dimensional nonlinear datasets. Consequently, this suggests that ScPace can be effectively applied to a diverse range of time-series ScRNA-seq datasets for the purpose of timestamp calibration.
\end{itemize}

The following passages are organized as follows. In Section \ref{related}, we introduced some related work. In Section \ref{ScPace}, our proposed method ScPace is described here. In Section \ref{results} the experimental design and experiment results of cross validation with different mislabeling rate and enhancement of supervised pseudotime analysis are shown. Moreover, We also included two case studies to illustrate the practical implications of using ScPace in real-world applications in Section \ref{case}, and the computational timing of ScPace in Section \ref{timing}. The final discussion are in Section \ref{discussion} and conclusion in Section \ref{conclusion}.
\section{Related Work}\label{related}
\subsection{Cell Type Annotation}
Cell type annotation in ScRNA-seq can be approached through both supervised and unsupervised methods, each offering unique advantages and challenges. Unsupervised methods, such as clustering algorithms\cite{clust}, have gained significant traction due to their ability to uncover novel cell populations without prior knowledge of cell identities, such as hierarchical clustering\cite{hie}, k-means clustering\cite{k_means}, and density-based approaches\cite{density}. Subsequently, these clusters can be characterized and annotated by integrating additional information, such as marker gene expression, functional gene sets, or reference atlases. In contrast, supervised methods for cell type annotation rely on labeled training datasets to inform the classification process. Techniques such as support vector machines\cite{scpred}, random forests\cite{singlecellnet}, and artificial neural networks\cite{ann_neural} can be employed to effectively learn the features that distinguish different cell types. Supervised approaches typically yield higher accuracy when reliable training datasets are available, facilitating the identification and characterization of specific cell types of interest.
\subsection{Pseudotime Analysis}
Pseudotime analysis is a computational method used primarily in time-series ScRNA-seq to infer the developmental trajectories of cells over time, and each cells are assigned a continues pseudotime value. Over the past decades, most unsupervised pseudotime analysis methods focused on both trajectory inference and pseudotime reconstruction, e.g., monocle\cite{monocle}, Seurat\cite{seurat}. Moreover, Tempora\cite{tempora} and IDTI\cite{idti} make use of the given timestamps for the inference of developmental trajectories. Methods like PseudoGA\cite{pseudoga}, TSCAN\cite{tscan} focus on assigning pseudotime ordering to each cells. However, unsupervised methods for assigning pseudotime often fails to correlate with the true timestamps. Therefore, to address this issue, supervised pseudotime analysis, Psupertime\cite{psupertime}, has been proposed to overcome this problem. By leveraging penalized ordinal regression, it could both handle timestamp automatic annotation and pseudotime ordering. Psupertime also facilitates the selection of genes that exhibit high correlation with temporal timestamps. Thus, combining trajectory inference methods and supervised pseudotime analysis could lead to a better interpretation of dynamic biological process.

However, in scenarios where multiple cells with varying timestamps are mislabeled, the efficacy of supervised pseudotime ordering may be compromised, potentially leading to suboptimal outcomes. This is because supervised pseudotime analysis heavily depend on the quality of the given timestamp labels to yield satisfactory performance. Therefore, the challenges arising from the reliance on ordinal timestamp labels in supervised pseudotime analyses are critical, as they can result in flawed interpretations of temporal dynamics. Although Psupertime has evaluated its performance in the context of label perturbations, it still fails to produce reliable results when confronted with noisy time-series ScRNA-seq data.
\subsection{ScReClassify}
ScReClassify marks a significant advancement in mislabeled cell type detection. This innovative semi-supervised learning framework addresses the challenge of mislabeled cells in ScRNA-seq datasets, offering a solution to refine cell type annotations generated by existing classification methods.

By leveraging dimensional reduction techniques such as PCA and applying a semi-supervised learning approach Adasampling\cite{Adasampling}, ScReClassify effectively identifies and reclassifies potentially mislabelled cells to their correct cell types. The framework has been validated using both simulated and real ScRNA-seq datasets across various tissues and biological systems, demonstrating its ability to accurately correct misclassified cells in ScRNA-seq data.

Despite its advantages, ScReclassify presents notable limitations. First, probability sampling methods frequently select less informative samples during each ensembles, which can lead to suboptimal results in timestamp automatic annotation and the enhancement of supervised pseudotime analysis. Second, ScReclassify relies on principle component analysis (PCA) for the dimensional reduction of the original data in all conditions, this approach could lead to information leakage and therefore degrade the performance on some time-series ScRNA-seq datasets.

\section{Methodology}\label{ScPace}
\subsection{ScPace}
\begin{figure}[h]   
\centering         
\includegraphics[scale=0.065]{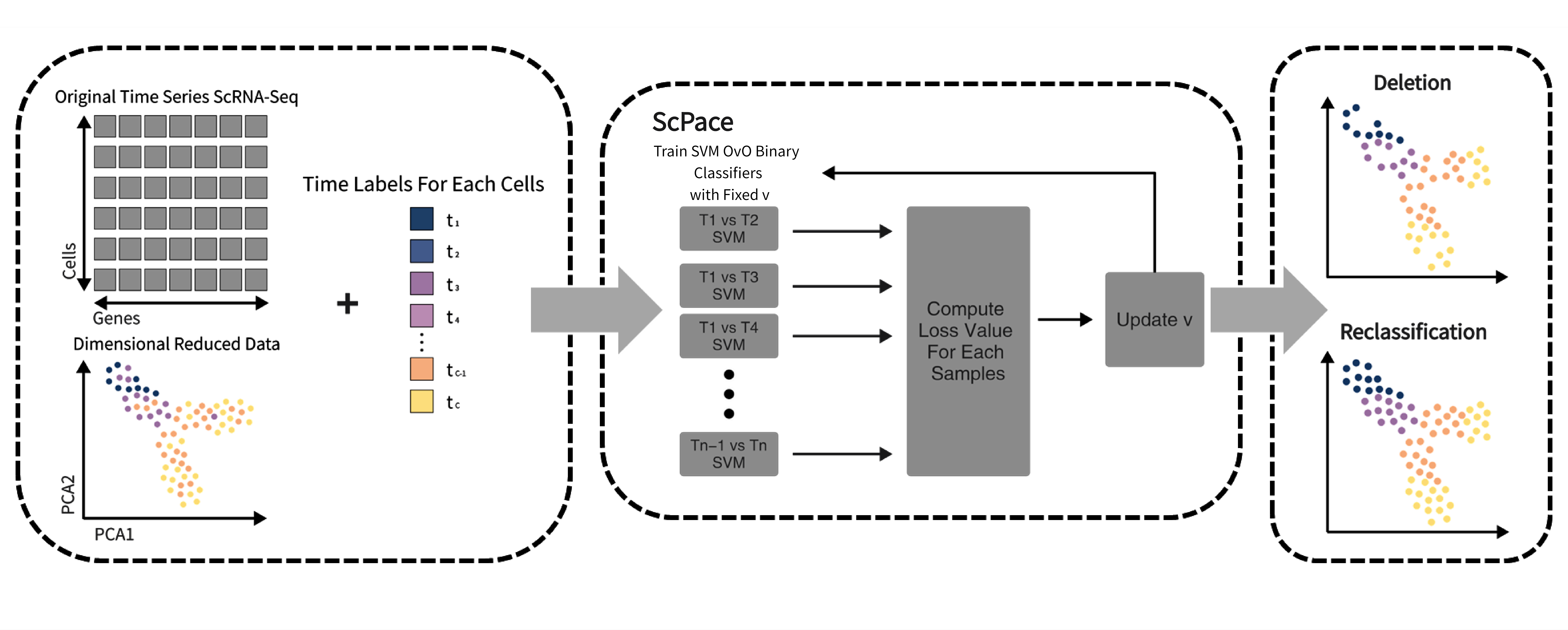}
\caption{Overall procedure of ScPace including data preprocessing, training ScPace and conduct two types of timestamp calibration.}
\label{overall}
\end{figure}

We developed ScPace by introducing a latent variable indicator designed to detect potential noisy labeled timestamps. The overall framework of ScPace is illustrated in Fig.(\ref{overall}). For our Base Classifier, we chose Support Vector Machines (SVM), which are well-suited for high-dimensional datasets. Although ScPace can be applied to other classifiers, such as neural networks, we opted not to experiment with them due to their propensity to overfit when analyzing time-series ScRNA-seq data. ScPace builds upon the concept of self-paced learning, but it differs in that traditional self-paced learning typically trains on all available data over predefined iterations. In contrast, ScPace focuses only on the subsets of data considered to be cleaned samples, allowing it to better handle noisy labeling. The joint optimization problem central to ScPace is addressed through a two-step iterative process, where the algorithm consistently refines its learning based on the identified clean samples over a specified number of iterations. This approach enhances the model's robustness and improves its ability to accurately classify cells in the presence of mislabeled data.

\subsubsection{Preprocessing of Dimensional Reduction}
Similar to ScReclassify, ScPace offers the option to conduct dimensional reduction on both training and testing sets, however, this preprocessing procedure is optional. Here, we implemented Principal Component Analysis (PCA) and Kernel Principal Component Analysis (KernelPCA)\cite{KernelPCA}. KernelPCA is an extension of PCA that introduces kernel mapping functions for nonlinear dimensionality reduction. Following the work of ScReclassify. we considered the original dimensions of the time-series ScRNA-seq datasets as $m \times n$. We determined $d$ as the reduced dimensions, where the number of principal components (PCs) is required to capture at least 70\% of the overall variability in the dataset. After obtaining $d$, the reduced dimensions are defined in Eq. (\ref{equa:001}), transforming the datasets into $m' \times n$. For kernelPCA, the dimensions are set to default $m^{'} = 20$. Note that we also used $m^{'} = 10$ in our experiments on both PCA and KenelPCA.
\begin{equation} 
	\label{equa:001}
 m^{'}=
    \left\{
    \begin{aligned}
    &10, d<10\\
    &d, 10<d<20 \\
    &20, d>20\\
    \end{aligned}
    \right.
\end{equation}
\subsubsection{Optimize One-Vs-One binary classifiers with latent variable}\label{step1}
In this section, we will introduce the methodology for training a single weighted One-vs-One binary Support Vector Machine (SVM) classifier with the introduced latent variable. This latent variable will be further optimized in subsequent sections. The One-vs-One approach allows us to construct multiple binary classifiers for each pair of classes, enabling effective discrimination even in multi-class scenarios. In binary cases, therefore $y=1$ are assigned to one classes and $y=-1$ are assigned to the distinct class. Given the reduced or original training sets $T = \{(x_{1},y_{1}),(x_{2},y_{2}),\dots,(x_{n},y_{n})\}$ where $x_{i} \in \{R^{m^{'}},R^{m}\},i = 1,2,\dots,n$, $y \in \{-1,+1 \}$, The standard soft-margin SVM classifier is being optimized with the primal problem equation:
\begin{align}\label{1}
\begin{aligned}\min_ {w, b, \zeta} & \frac{1}{2} w^T w + C \sum_{i=1}^{n} \zeta_i\\
\begin{split}
\textrm {s.t. } & y_i (w^T \phi (x_i) + b) \geq 1 - \zeta_i,\\
& \zeta_i \geq 0, i=1, ..., n
\end{split}
\end{aligned}
\end{align}

where $w,b$ is the optimal parameter and the hyperparameter $C$ controls the penalty strength of the misclassified points. Eq.(\ref{1}) could be rewritten with the given latent variables $v_{i} \in \{0,1\}$, $i = 1,2,\dots, n$ which are shown in Eq.(\ref{2}), If $t = 1 $, then it indicates the initial optimization phase. During this stage, the model parameters are set to their starting values, and the latent variables are constants, which means that in Section(\ref{step1}) the optimization is not being done on the latent variable $v_{i}$. If $ \Vert v \Vert_1=n$ , meaning that all samples are included for training. However, when $t \neq 1$, the L1-norm on latent variable $ \Vert v \Vert_1$ might not equal to $n$, meaning that some samples are not included for training.
\begin{align}\label{2}
\begin{aligned}\min_ {w, b, \zeta} & \frac{1}{2} w^T w + C \sum_{i=1}^{n} v_{i}\zeta_i\\
\begin{split}
\textrm {s.t. } & v_{i}[y_i (w^T \phi (x_i) + b)] \geq v_{i}(1 - \zeta_i),\\
& \zeta_i \geq 0, i=1, ..., n
\end{split}
\end{aligned}
\end{align}

For dealing with data imbalancing, we introduce different penalty parameter for each classes in terms of class ratio\cite{weight_svm}. Therefore, given two classes, Eq.(\ref{2}) could be rewritten as:
\begin{align}\label{3}
\begin{aligned}\min_ {w, b, \zeta} & \frac{1}{2} w^T w + C_{1} \sum_{y_{i}=1} v_{i}\zeta_i +C_{2}\sum_{y_{i}=-1} v_{i}\zeta_{i}\\
\begin{split}
\textrm {s.t. } & v_{i}[y_i (w^T \phi (x_i) + b)] \geq v_{i}(1 - \zeta_i),\\
& \zeta_i \geq 0, i=1, ..., n
\end{split}
\end{aligned}
\end{align}
Given the count vector for each classes $B =(\displaystyle \sum_{y=1}1,\displaystyle\sum_{y=2}1,\dots,\sum_{y=N}1)$ , penalization hyperparameter $C$, and the number of samples $n$, the penalization for each class is being calculated in Eq.(\ref{weight_count}):
\begin{align}\label{weight_count}
    c =(\frac{C\times n}{m\displaystyle\sum_{y=1}1},\dots,\frac{C\times n}{m\displaystyle\sum_{y=N}1})
\end{align}
However, if the two classes are balanced, then $C_{1}=C_{2}=C$, which means that the modified Eq.(\ref{3}) with different penalized parameter for each class would then equivalent to the previous format Eq.(\ref{2}).

Moreover, The rewritten equation could be further transformed into hinge loss format which is equivalent to optimizing the primal problem shown in Eq.(\ref{svm_new}).
\begin{align}\label{svm_new}
\begin{split}
    \min_{w,b}&\frac{1}{2}w^{T}w+C_{1}\sum_{y_{i}=1}v_{i}[1-y_{i}(<w,\phi (x_{i})>+b)]_{+} +\\ 
    &C_{2}\sum_{y_{i}=-1}v_{i}[1-y_{i}(<w,\phi (x_{i})>+b)]_{+}
\end{split}
\end{align}

However, since we introduced kernel computation to solve non-linear problems, therefore, ScPace solves the following dual problem of SVM with the latent variables $v_{i}$ to compute the dual parameter. which is equivalent to Eq.(\ref{svm_new}). Consider constructing the Lagrangian form of the primal problem with the given latent variables $v_{i}$:
\begin{align}\label{dual_lagaga}
\begin{aligned}
&L(w,b,\zeta,\alpha,\mu,v) = \frac{1}{2}w^{T}w+C_{1}\sum_{y_{i}=1}v_{i}\zeta_{i}\\
&+C_{2}\sum_{y=-1}v_{i}\zeta_{i}-\sum_{i=1}^{n}\alpha_{i}v_{i}[y_{i}(w^{T}\phi(x_{i})+b)-1+\zeta_{i}]\\
&-\sum_{i=1}^{n}\mu_{i}v_{i}\zeta_{i}\\
\begin{split}
\textrm  {s.t. } v_{i}\alpha_{i}\geq 0, v_{i}\mu_{i}\geq 0
\end{split}
\end{aligned}
\end{align}
Since the dual problem is a maximizing optimization problem on the minima of Lagrangian functions. Therefore, we must seek partial derivatives of $w$,$b$,$\zeta_{i}$ of Eq.(\ref{dual_lagaga}).
\begin{align}
    \frac{\partial L(w,b,\zeta,\alpha,\mu,v)}{\partial w} &= w-\sum^{n}_{i=1}\alpha_{i}v_{i}y_{i}x_{i}=0 \label{div1}\\ 
    \frac{\partial L(w,b,\zeta,\alpha,\mu,v)}{\partial b} &= -\sum_{i=1}^{n}\alpha_{i}v_{i}y_{i}=0 \label{div2}\\
    \frac{\partial L(w,b,\zeta,\alpha,\mu,v)}{\partial \zeta_{i}} & =
    \begin{cases}\label{div3}
    v_{i}(C_{1}-\alpha_{i}-\mu_{i})=0,y_{i}=1 \\
    v_{i}(C_{2}-\alpha_{i}-\mu_{i})=0,y_{i}=-1 \\
    \end{cases}
\end{align}
Then Eq.(\ref{div1})(\ref{div2})(\ref{div3}) are taken into Eq.(\ref{dual_lagaga}) for computing the minimal result on $w$,$b$,$\zeta$ as shown in the following equation:
\begin{align}\label{lagaga_min}
\begin{split}
    &\min_{w,b,\zeta} L(w,b,\zeta,\alpha,\mu,v) = \min_{w,b,\zeta} -\frac{1}{2}\sum^{n}_{i=1}\sum^{n}_{j=1}\alpha_{i}\alpha_{j}v_{i}v_{j}y_{i}y_{j}K(x_{i},x_{j})+\sum^{n}_{i=1}\alpha_{i}v_{i},
\end{split}
\end{align}
where $K(x_{i},x_{j})$ is the kernel computation. For ScPace, We only validated our experiments using RBF Kernel functions where $K(x_{i},x_{j}) = \exp{(-\gamma {\Vert x_{i}-x_{j}\Vert}^{2})}$. In order to obtain the optimal $w$ and $b$ in dual space, it requires to maximize Eq.(\ref{lagaga_min}) on parameter $\alpha$. Therefore, we obtain the following optimization problem:
\begin{align}
\begin{aligned}\max_{\alpha} \min_{w,b,\zeta} L(w,b,\zeta,\alpha,\mu,v) &= \max_{\alpha} \min_{w,b,\zeta}-\frac{1}{2}\sum^{n}_{i=1}\sum^{n}_{j=1}\alpha_{i}\alpha_{j}v_{i}v_{j}y_{i}y_{j}K(x_{i},x_{j})+\sum^{n}_{i=1}\alpha_{i}v_{i}\\
\begin{split}
\textrm {s.t. } & 0 \leq v_{i}\alpha_{i}\leq C_{1}, y_{i}=1\\
& 0 \leq v_{i}\alpha_{i}\leq C_{2}, y_{i}=-1.\\
\end{split}
\end{aligned}
\end{align}
After solving the optimization problem, we could then obtain the optimal $\alpha$ in terms of KKT condition. Consider we obtained the optimal solution $\alpha^{*}=(\alpha^{*}_{1},\alpha^{*}_{2},\dots,\alpha^{*}_{n})$:
\begin{align}
    w^{*} &= \sum^{n}_{i=1}\alpha_{i}^{*}y_{i}v_{i}x_{i}\\
    b^{*} &= y_{j}-\sum_{i=1}^{n}\alpha^{*}_{i}v_{i}y_{i}K(x_{i},x_{j})
\end{align}
note that the optimal parameter $b^{*}$ is obtained from support vectors and optimal parameter $w^{*}$ is obtained from Eq.(\ref{div1}). Therefore, we could obtain the decision boundary corresponding to its decision function with the latent variables after obtaining the optimal solution:
\begin{align}\label{decision}
    f(x) = sign(\sum^{n}_{i}\alpha^{*}_{i}v_{i}y_{i}K(x,x_{i})+b^{*})
\end{align}
\subsubsection{Optimize latent variable with optimized parameters $w$ and $b$}\label{step2}
This is the core structure of ScPace where we introduce a latent variable indicator $v$, this enables ScPace to identify which samples are considered as potential noisy labeled samples and therefore remove them for training. The mathematical forms shown in Eq.(\ref{selection}) of removing them is by assigning 0 to the samples which are higher than the loss value and 1 to the samples which are lower than the given threshold $\lambda$.
\begin{align}\label{selection}
v_{i}=
    \begin{cases}
    1,L_{i}(y,W)<\lambda \\
    0,otherwise\\
\end{cases}
\end{align}
For multi-class problem, SVM initially adopts the ove-vs-one training strategy. After Computing the parameter $w$ and $b$ in terms of the dual coefficients, we therefore could obtain $\frac{N(N-1)}{2}$ Decision Values for each samples according to Eq.(\ref{decision}). Here we transformed the decision value shape into $N$ dimensions using a transformed confidence strategy.

\begin{algorithm}
    \label{a1}
    \SetKwInOut{KIN}{Input}
    \SetKwInOut{KOUT}{Output}
    \caption{Transforming OvO Decision Function Output into OvR Output}
    \KIN{Prediction matrix for each samples $P$,
    Decision Value Matrix for each Samples $D$}
    \KOUT{Vote Matrix $\hat{V}$, Sum confidence Matrix $S$}

    \textbf{Initialize:} U, S, V,$k=0$\;
    
    \For{$i \leftarrow 0$ \KwTo $N$}{
        \For{$j \leftarrow i+1$ \KwTo $N$}{
        S[:, i] -= D[:, k];
        
        S[:, j] += D[:, k];
            
        $\hat{V}[P[:, k] == 0, i] += 1$;
        
        $\hat{V}[P[:, k] == 1, j] += 1$;
            
        k += 1;
        
        }
    }
    \Return{ $\hat{V}$, $S$}
\end{algorithm}

In terms of Algorithm \ref{a1}, we therefore could obtain the voting matrix $\hat{V}$ and the sum confidence matrix $S$, we then monotonically transform $S$ according to the following Equation:
\begin{align}\label{transform}
    \hat{a_{ij}} = \frac{a_{ij}}{3(\left| a_{ij} \right|+1)}
\end{align}
Where $\hat{a_{ij}} \in \hat{S}$, $a_{ij} \in S$, ($i = 1,2,\dots,n$, $j = 1,2,\dots,N$). The final decision value transformed into One-vs-Rest Shape are therefore the sum of voting matrix and transformed sum confidence matrix $\hat{S}$.
\begin{align}\label{final}
    W = \hat{S}+\hat{V}
\end{align}
After we obtained the decision value in $N$ dimensional shape for each samples according to Eq.(\ref{final}), we used multi-class hinge loss proposed by Crammer \& singer\cite{cramer} for computing the loss value for each samples. Therefore, considering $n$ samples and $W_{i,y_{i}}$ is the predicted decision value for the corresponding label $y_{i}$ of the i-th sample, and $\hat{W_{i,y_{i}}}=max\{W_{i,y_{i}}|y_{j} \neq y_{i}\}$, the loss value for each samples are shown in the following equation:
\begin{align}\label{loss}
    L_{i}(y,W) = \max\{ 1+W_{i,y_{i}}-\hat{W_{i,y_{i}}},0\}
\end{align}
After Computing the loss value for each samples with Eq.(\ref{loss}), We could optimize the latent variables with the following modified version to consider which samples are clean or noisy:
\begin{align}\label{spl}
    &min_{v} \sum^{n}_{i=1}v_{i}L_{i}(y,W) -\lambda \sum^{n}_{i=1}v_{i} \\
    &min_{v} \sum^{n}_{i=1}v_{i}(L_{i}(y,W) -\lambda)\label{spl2}
\end{align}
where $\lambda$ is the given threshold. Eq.(\ref{spl2}) is a modified version of Eq.(\ref{spl}), Since $v_{i} \in \{0,1\}$. samples which obtain loss value $L_{i}(y,W)$ greater than the threshold $\lambda$ would not contribute in minimizing Eq.(\ref{spl2}).
\subsubsection{Overall Algorithm}
The aforementioned two optimization procedures are conceptualized as a singular iterative step, which can be repeated multiple times to attain the desired outcomes. An additional noteworthy characteristic of ScPace is the incremental adjustment of the parameter $C$ following each iteration, where $C$ is augmented at a specified rate $p$. This addresses the issue of model underfitting that may arise subsequent to the removal of samples, while concurrently mitigating the risk of excessive sample elimination.

\begin{algorithm}
    \SetKwInOut{KIN}{Input}
    \SetKwInOut{KOUT}{Output}
    \caption{Timestamps Calibration Using ScPace}
    \KIN{Training Sets $T = \{(x_{1},y_{1}),(x_{2},y_{2}),\dots,(x_{n},y_{n})\}$, Threshold Hyperparameter $\lambda$, Regularization Parameter $C$, Calibration Iteration L, Penalization Vector $c$}
    \KOUT{Latent Variable Vector $V$, Model Parameter vector $w$,$v$}
    
    initialization $w$,$b$ and $V$\;

    \If {Dimensional Reduction}{
        Conduct PCA or KernelPCA on Original Datasets
        }
    
    \For{$n \leq L$}{
        $n = n+1$\;
        
        $C = C + p$\;
        \If {Class Not Balanced}{
        Recalculate $c$ Penalization Vector\;

       c = c*C \;
        }
        
        \While{not converged}{
            Train One-vs-One Classifier With Updated latent variable Vector $V$\;
        }
        Transform decision function value into $N$ dimension by Algorithm 1 and Eq.(\ref{transform})
        
        Compute Loss Value for each samples by Eq.(\ref{loss})

        Update Latent Variable Vector $V$ by Eq.(\ref{spl2})
    }
    \Return $V$,$\{w,b\}=\{w^{*},b^{*}\}$\;
\end{algorithm}
\newpage
\section{Experiments}\label{results}
\subsection{Data Preparation}\label{data}
\begin{figure}[h]   
\centering         
\includegraphics[scale=0.065]{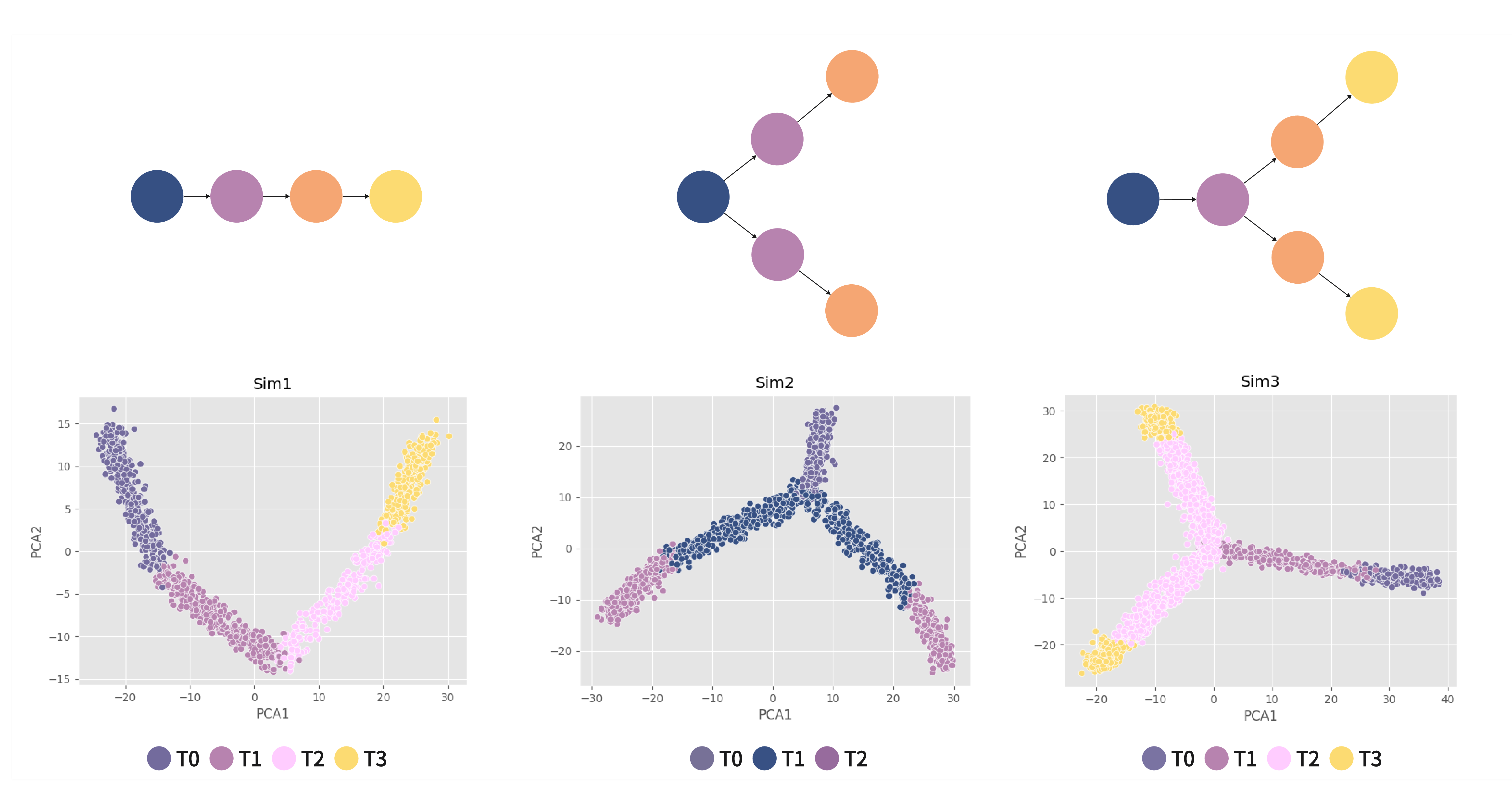}
\caption{Visualization of the simulated datasets simulated by Splatter, Sim1 and Sim2 are class balanced while Sim3 is imbalanced.}
\label{sim_present}
\end{figure}

We used the R package Splatter\cite{splatter} to simulate two types of trajectories: a linear path and a bifurcation path\cite{peihong}. For the linear path, designated as Sim1, we simulated 1,200 cells, each characterized by 6,000 genes. Each cell state corresponds to a single time point, resulting in four different time points, each with ordinal labels ($T_{1}, T_{2}, T_{3}, T_{4}$). Next, we simulated two bifurcation paths, one balanced and one imbalanced. The balanced data, referred to as Sim2, consists of 1,500 cells and 8,000 genes, with three time points ($T_{1}, T_{2}, T_{3}$). The imbalanced data, referred to as Sim3, includes 2,000 cells and 10,000 genes, featuring four time points ($T_{1}, T_{2}, T_{3}, T_{4}$). For the balanced bifurcation path in Sim2, the cell state prior to the split into two branches is defined as a single time point. Here, State 1 (S1) is considered as $T_{1}$. In Sim3, the same labeling is applied, with S1 as $T_{1}$ and State 2 (S2) as $T_{2}$. After the split in Sim2, both State 3 and State 4 are assigned to $T_{2}$, while States 5 and 6 are assigned to $T_{3}$. In Sim3, States 3 and 4 are considered as $T_{3}$, and States 5 and 6 correspond to $T_{4}$. We also simulated dropouts, which is a common feature in ScRNA-seq datasets. Subsequently, we conducted gene filtering on the simulated datasets to enhance the data quality for further analysis.The visualization of the simulated datasets are shown in Fig.(\ref{sim_present}). 

For real time-series ScRNA-seq datasets, all datasets were publicly downloaded from the US National Library of Medicine, National Institutes of Health. For GSE90047 and GSE98664\cite{gse98664}, we first log-normalized them using the Python Scanpy\cite{scanpy} package and removed genes that were expressed in fewer than 10 cells. Cells with low counts were also excluded if the total number of genes expressed in a cell was lower than 200. However, given that the original dataset for GSE98664 comprised 157,717 genes, we subsequently selected only the highly variable genes for further analysis using Scanpy. In the case of GSE67310\cite{gse67310}, the dataset was already normalized upon download, therefore, our preprocessing focused on gene and cell filtering only. For dataset E-MTAB-3929, low quality cells and genes are first removed and then we only selected highly variable genes for further analysis. Additional statistical information regarding both the simulated and real time-series ScRNA-seq datasets is presented in Table \ref{stats}.
\begin{table}\label{simulate}
\centering
\setlength{\tabcolsep}{7mm}
\begin{tabular}{ccccccc}
\toprule
\textbf{} & Datasets &Cells & Genes & Classes & Class Balanced\\
\midrule                                                                                                                                                                  
\textbf{Simulated}&Sim1 & 1200 & 6000 & 4& T\\
\textbf{}&Sim2 & 1500 & 8000 & 3& T\\
\textbf{}&Sim3 & 2000 & 10000 & 4& F\\

\textbf{Real}&GSE90047 & 439 & 21361 & 7 & T \\
\textbf{} & GSE67310 & 405 & 13202& 5  & T\\
\textbf{} & GSE98664 & 421 & 22756& 5  & T\\
\textbf{} & E-MTAB-3929 & 1529 & 10235& 5  & F\\
\bottomrule
\end{tabular}
\caption{Statistical results on Simulated Datasets and real time-series ScRNA-seq datasets, each rows correspond to different datasets, Sim1 and Sim2 are class balanced datasets with different trajectory, linear and bifurcation, and Sim3 correspond to imbalanced dataset on bifurcation path. Genes and cells are listed in terms of the preprocessed datasets.}
\label{stats}
\end{table}
\subsection{Cross validation with different mislabeling rate}
\subsubsection{Experimental Setup}
\begin{figure}[h]   
\centering         
\includegraphics[scale=0.065]{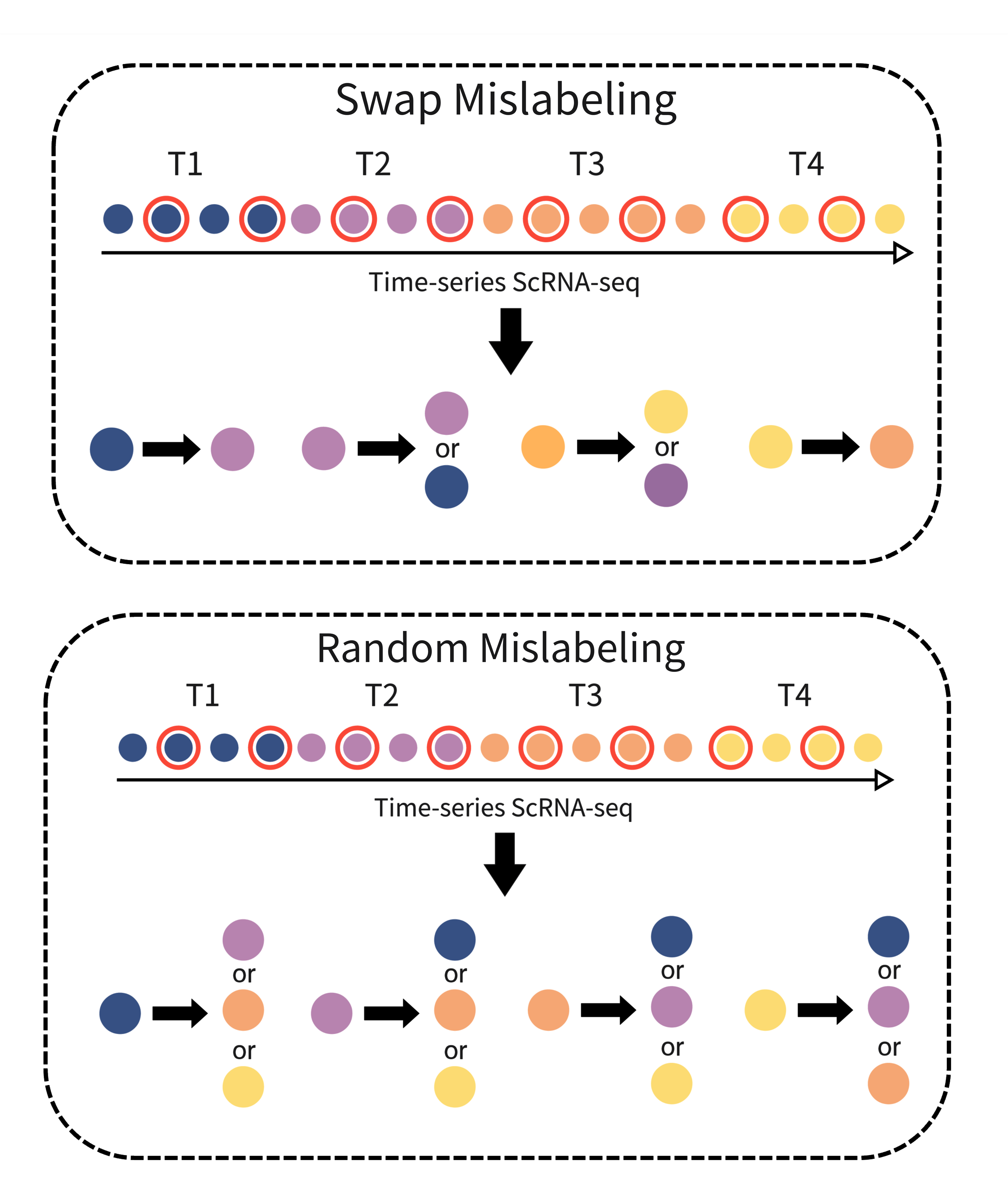}
\caption{Demonstration of Swap Mislabeling and Random Mislabeling}
\label{mislabeling}
\end{figure}

We conducted stratified k-fold(k=5) cross validation on the simulated datasets and real time-series ScRNA-seq datasets. k-fold stratified cross validation is a variant of the k-fold cross-validation method that aims to ensure that the class distribution in the dataset is maintained across all folds. The term "stratified" means that the division is done in a way that preserves the original class distribution of the data. This is particularly useful when dealing with imbalanced datasets, where the number of samples in different classes varies significantly. In our cross-validation experiments, the training sets are artificially mislabeled and the holdout testing sets are cleaned without artificial mislabeling. When assigning the artificial mislabeling on the training sets, there are two types of mislabeling which are shown in Fig.(\ref{mislabeling}).

\begin{itemize}
    \item The first type of mislabeling we consider is known as swap mislabeling. In this method, cells are randomly selected from the training sets based on a specified mislabeling rate (e.g., 0.1, 0.2, 0.3, or 0.4). Each cell is assigned a decision value of either $0$ or $1$. In this case, $1$ indicates that the cell will be mislabeled, while $0$ signifies that the cell remains clean (i.e., correctly labeled). For the cells marked for mislabeling, their original labels are swapped with labels from neighboring time points.  For cells labeled at intermediate time points $T_{m}$ where $T_{1}<T_{m}<T_{n}$, the selected cells are mislabeled as either $T_{m-1}$ or $T_{m+1}$. Specially, cells labeled as $T_{1}$ would be mislabeled as $T_{2}$, for cells labeled with  $T_{n}$, they would instead be mislabeled as $T_{n-1}$. This structured approach to mislabeling helps to maintain some degree of temporal coherence in the dataset, as the labels are swapped only with those from neighboring time points. This mimics potential real-world scenarios where labels may become inaccurate but remain somewhat related to nearby observations in time-series ScRNA-seq datasets.
    \item The second type of mislabeling we examined is random mislabeling. In this scenario, cells are selected randomly from the training sets based on the specified mislabeling rate. Unlike swap mislabeling, where labels are exchanged between neighboring cells, random mislabeling involves assigning random entirely new, distinct labels to the selected cells, which differ from their original labels. This approach introduces a higher level of uncertainty, as the mislabeling occurs independently for each selected cell rather than being influenced by nearby or related cells. As a result, random mislabeling tends to create a more challenging environment for timestamp automatic annotation and supervised pseudotime analysis, making it crucial to assess their performance under this condition.
\end{itemize}

we compared our proposed method against several traditional machine learning classifiers that do not specifically account for noise in the data. For this comparison, we included two tree-based classifiers: Decision Tree and Random Forest. Additionally, we incorporated a deep learning approach using a fully connected neural network classifier. It is important to note that ScReclassify is not originally designed for predicting labels for new cells. To address this limitation, we extended the methodology by performing dimensional reduction on the testing set, using the parameters determined from the training set. Moreover, we have included fully connected neural network based confident learning\cite{2021confident} using CleanLab for timestamp calibration.

For the setting of the model hyperparameters, we utilized the grid search method to systematically tune the hyperparameters of the models, thereby enabling a comprehensive comparison of the effectiveness of each classifier. The grid search method is an exhaustive approach for hyperparameter tuning in machine learning models, involving a thorough search through a manually specified subset of the hyperparameter space to identify the optimal combination that yields the best model performance on a validation set. Particularly, special attention was given while conducting the cross-validation experiments on the fully connected neural network, as it is prone to extreme overfitting. The parameters set for the grid search are presented in Table \ref{param_cross}:

\begin{table}
\centering
\setlength{\tabcolsep}{0.25mm}
\begin{tabular}{cccc}
\toprule
Classifiers & Parameter & Description & Values\\
\midrule                                                                                                                                                                  
\textbf{ScPace} & \textbf{C} & The regularization Parameter & \{0.1,1,10\}\\
\textbf{} & \textbf{p} & The C-Growing Parameter & \{0.1,0.5,1\}\\
\textbf{} & \textbf{Iter} & Number of iterations to perform ScPace & \{5,10\}\\
\textbf{} & \textbf{$\lambda$} & Threshold for updating the latent variables $v$ & \{0.1,0.5,1,2\}\\

\textbf{ScReclassify} & \textbf{L} & Number of ensembles & \{3,5,10\}\\
\textbf{} & \textbf{$\alpha\%$} & Percentage of samples to select at each iteration & \{0.5,0.8,1\}\\
\textbf{} & \textbf{Iter} & Number of iterations to perform Adasampling & \{3,5,10\}\\
\textbf{} & \textbf{Balance} & Logical flag if the cell types are balanced & \{0,1\}\\

\textbf{SVM} & \textbf{C} & The regularization Parameter & \{0.1,1,10\}\\

\textbf{Random Forest} & \textbf{MD} & The maximum depth of the tree & \{10,20,50,100\}\\

\textbf{Decision Tree} & \textbf{MD} & The maximum depth of the tree & \{10,20,50,100\}\\

\textbf{Neural Network} & \textbf{(i,j,k)} & The hidden layer sizes of the network& \{(50,50,50),(50,100,50)\}\\

\textbf{} & \textbf{$\rho(x)$} & The activation function of each neurons & \{$\max{(0,x)}$,$\frac{e^{x} - e^{-x}}{e^{x} + e^{-x}}$,$x$\}\\

\textbf{CL} & \textbf{(i,j,k)} & The hidden layer sizes of the network& \{(50,50,50),(50,50,100)\}\\

\textbf{} & \textbf{$\rho(x)$} & The activation function of each neurons & \{$\max{(0,x)}$\}\\
\bottomrule
\end{tabular}
\caption{Hyperparameters used for grid search tuning while conducting the cross validation experiments}
\label{param_cross}
\end{table}
We utilized several metrics to assess the performance of the cross-validation experiments on both balanced and imbalanced datasets. These metrics were chosen to evaluate the overall accuracy of our proposed method. It is important to note that during the cross-validation experiments on the imbalanced datasets, particularly Sim3, the models tended to learn biases toward the majority classes. This bias could result in a high overall accuracy score while reflecting low accuracy on minority classes.

To address this issue and provide a more comprehensive assessment of performance, we applied a balanced average score for the imbalanced datasets. This metric computes the average recall score obtained for each class\cite{macrof1}, as shown in Eq. (\ref{final_metric}). By employing this approach, we aimed to ensure that the evaluation reflects the model's true performance across all classes, particularly the minority ones. 
\begin{align}
    Acc(y_{i},\hat{y_{i}}) &= \frac{1}{n}\sum^{n-1}_{i=0}1(y_{i}=\hat{y_{i}})
\end{align}
Here we define any classifier $f : D \to L$ where $D = \{X_{1},X_{2},\dots,X_{n}\}$ correspond to training vector datasets and $L = \{0,1,2,\dots,m\}$. and cartesian product $S \subseteq D \times L =\{<x,y>|x \in D,y\in L\}$ . let $M \in \mathbb{N}^{m\times m}$ be the confusion Matrix and $M_{ij} = \left |\{ s \in S| f(s_{x}) = i \land s_{y}=j\}\right |$. Therefore, we could obtain the recall(R) score for each class $i = 0,1,2,\dots,m$ and the balanced accuracy $BA$ score for each samples:
\begin{align}\label{metrics}
    &R_{i} = \frac{M_{ii}}{\sum^{n}_{k=1}M_{ki}}\\
    &BA= \frac{1}{m}\sum_{i=0}^{m}R_{i}\label{final_metric}
\end{align}
Moreover, we also assessed the correction ability of ScPace, ScReclassify and CL using Eq.(\ref{correction}). This metric assess the amount of mislabeled samples corrected compared with the number of all mislabeling samples. Given mislabeling rate $\rho$, the artificial mislabeled samples $\{(x_{1},y_{1}),\dots,(x_{\rho n},y_{\rho n})\}$, the cleaned samples $\{(x_{1},y_{1}^{*}),\dots,(x_{\rho n},y_{\rho n}^{*})\}$ and the reclassified samples $\{(x_{1},\hat{y_{1}}),\dots,(x_{\rho n},\hat{y_{\rho n}})\}$:
\begin{align}
    C = \frac{\sum_{i=1}^{\rho n}1(y^{*}_{i}=\hat{y_{i}})}{\rho n}\label{correction}
\end{align}
\subsubsection{Experimental Results}
\begin{figure}[h]   
\centering         
\includegraphics[scale=0.075]{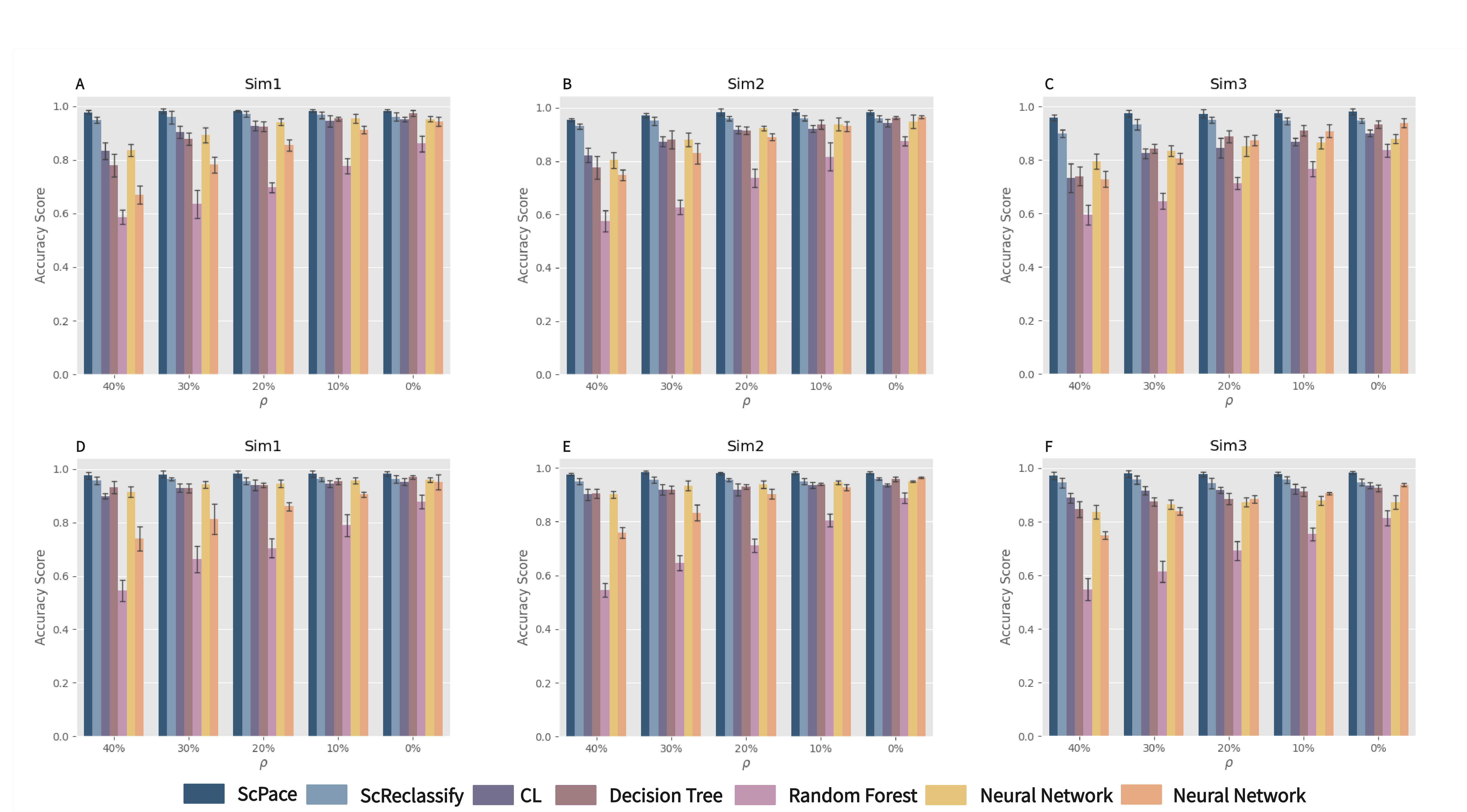}
\caption{Cross validation results with different mislabeling rate on simulated datasets (\textbf{A-C}) cross validation results with swap mislabeling (\textbf{D-F}) cross validation results with random mislabeling}
\label{scpace_cross}
\end{figure}

For both types of mislabeling, our detailed results shown in Fig.(\ref{scpace_cross}) indicate that on the balanced linear path Sim1, ScPace achieved an average accuracy score of $97.33\%$ at a mislabeling rate of $40\%$ for swap mislabeling and an average accuracy score of $98.06\%$ for random mislabeling. These results represent the highest accuracy scores observed at the maximum mislabeling rate 40\%, and ScPace outperformed traditional machine learning classifiers, including the random forest and neural network classifiers. Additionally, ScPace maintained superior performance across varying mislabeling rates.

Interestingly, while other classifiers exhibited notable sensitivity to mislabeling rates, ScPace demonstrated remarkable resilience. The model consistently achieved accuracy above $90\%$ regardless of the mislabeling rate. To quantitatively compare the significant differences in cross-validation results between ScPace and traditional machine learning classifiers, we performed the Mann-Whitney Wilcoxon Tests. The results indicated that the differences were statistically significant ($P < 0.01$) at the $40\%$ mislabeling rate. Furthermore, we observed that the significance of these differences diminished as the mislabeling rate decreased.

In a similar manner, on bifurcation path Sim2, ScPace achieved an average accuracy score of $95.47\%$ at a $40\%$ mislabeling rate for swap mislabeling and an average accuracy score of $97.67\%$ for random mislabeling. Again, ScPace outperformed other classifiers across different mislabeling rates. These consistent results underscore the robust performance of ScPace in the presence of mislabeling.

To further demonstrate the robustness of our method on noisy imbalanced datasets, we conducted cross-validation experiments on the imbalanced bifurcation simulated dataset. The experimental results reveal that ScPace outperformed other classifiers, achieving a balanced accuracy of $95.84\%$ at a $40\%$ mislabeling rate for swap mislabeling and $97.12\%$ for random mislabeling. The differences in performance were statistically significant, as determined by the Mann-Whitney Wilcoxon Tests ($P < 0.05$). These results suggest that ScPace is not only effective in balanced scenarios but also robust in the presence of noise and imbalance situations, making it applicable to a wide range of datasets.

\begin{figure}[h]   
\centering         
\includegraphics[scale=0.065]{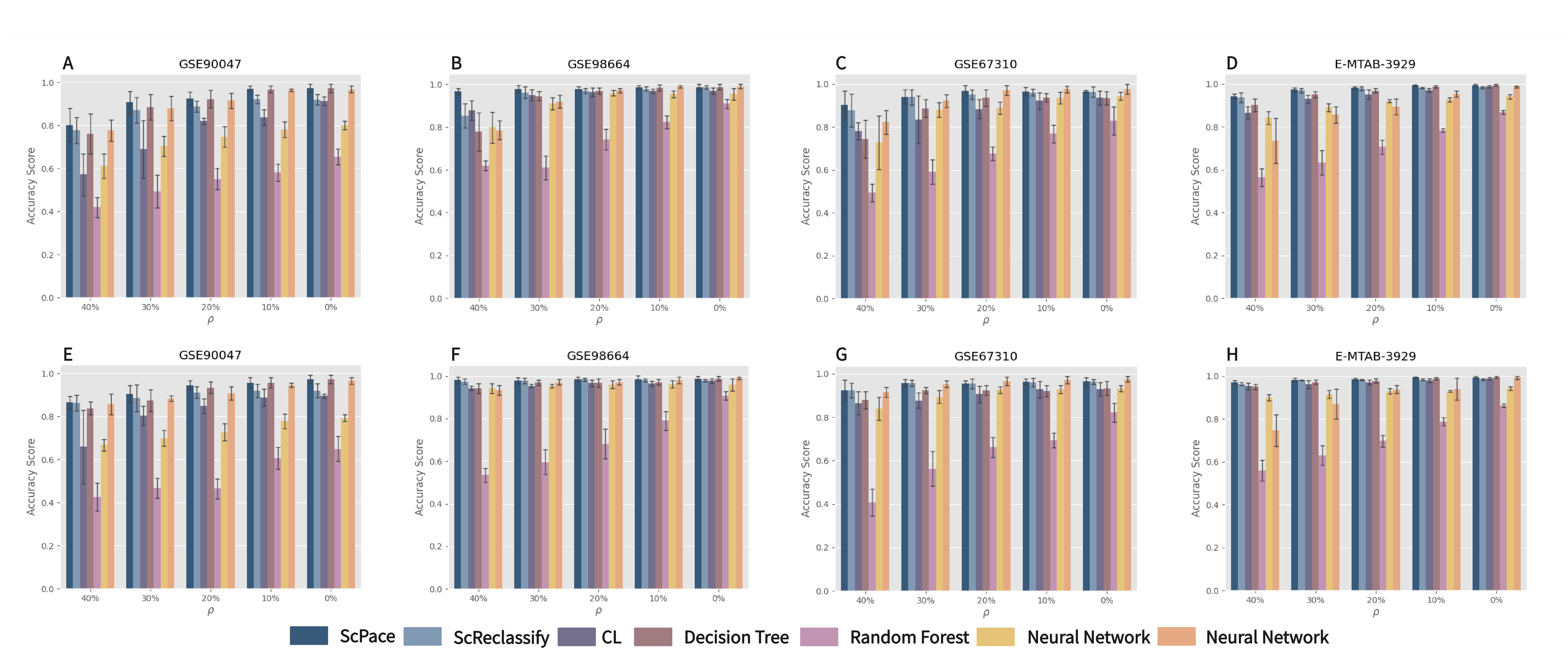}
\caption{Cross validation results with different mislabeling rate on real time-series ScRNA-seq datasets (A-D) cross validation results with swap mislabeling (E-H) cross validation results with random mislabeling}
\label{real_cross}
\end{figure}

On real time-series ScRNA-seq datasets, we also introduced two types of mislabeling for cross-validation experiments. The experimental results shown in Fig.(\ref{real_cross}) that on ScPace performed better than ScReclassify and traditional machine learning classifiers on most occasions, even when mislabeling rate reaches $40\%$. Interestingly, we found out that when mislabeling rate is under 40\%, the performance of SVM tends to perform better than ScReclassify. Interestingly, we found out that when there is no artificial mislabeling, all the classifier performed above accuracy score $95\%$, which indicates that the label noises in the testing sets didn't effect the overall performance, practically because the timestamps of most cells in the original are well labeled.

\begin{figure}[h]   
\centering         
\includegraphics[scale=0.065]{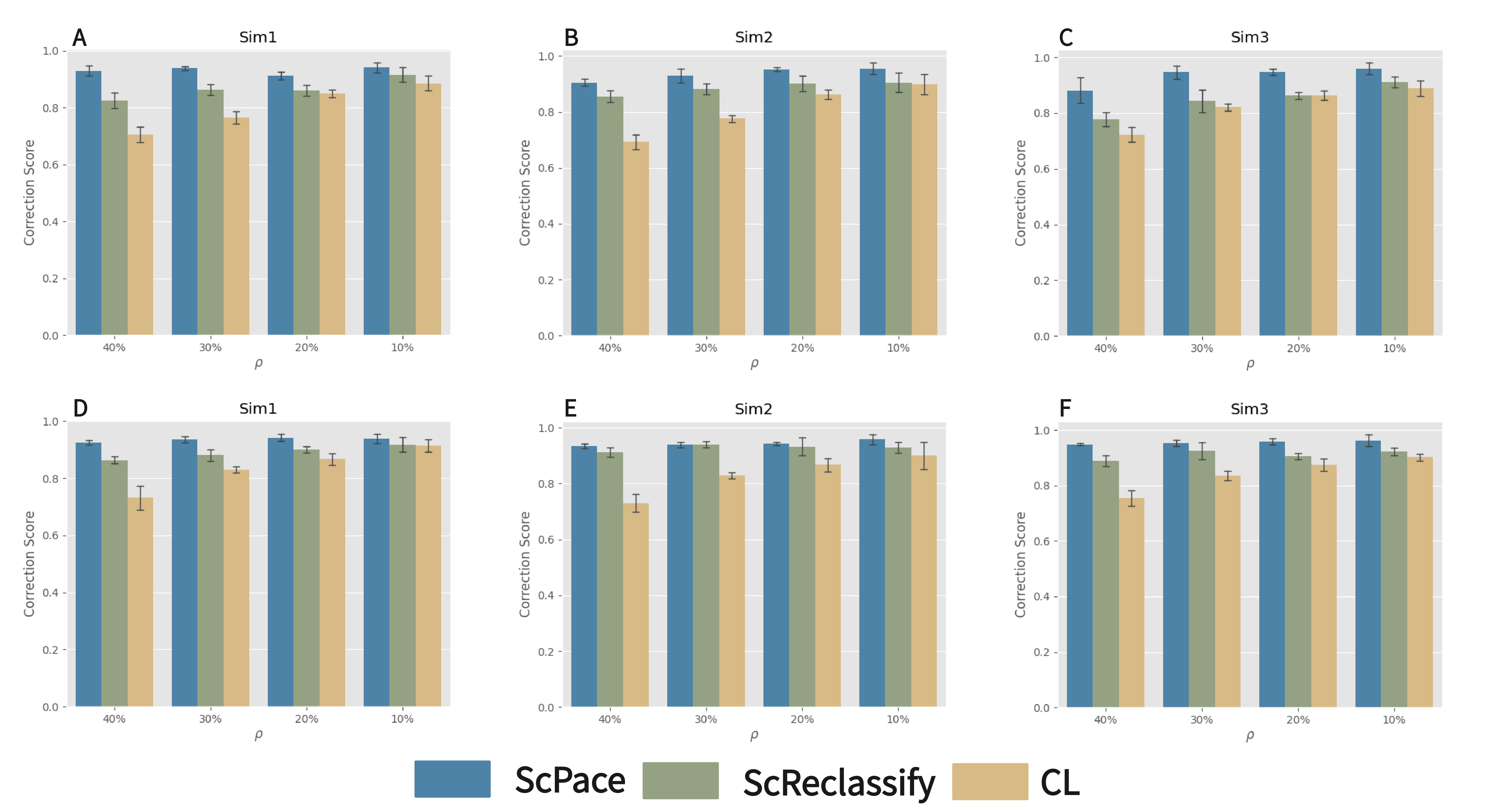}
\caption{Correction score of cross validation with different mislabeling rate on simulated datasets (A-C) correction score results with swap mislabeling (D-F) correction score results with random mislabeling}
\label{cor_score_sim}
\end{figure}

\begin{figure}[h]   
\centering         
\includegraphics[scale=0.065]{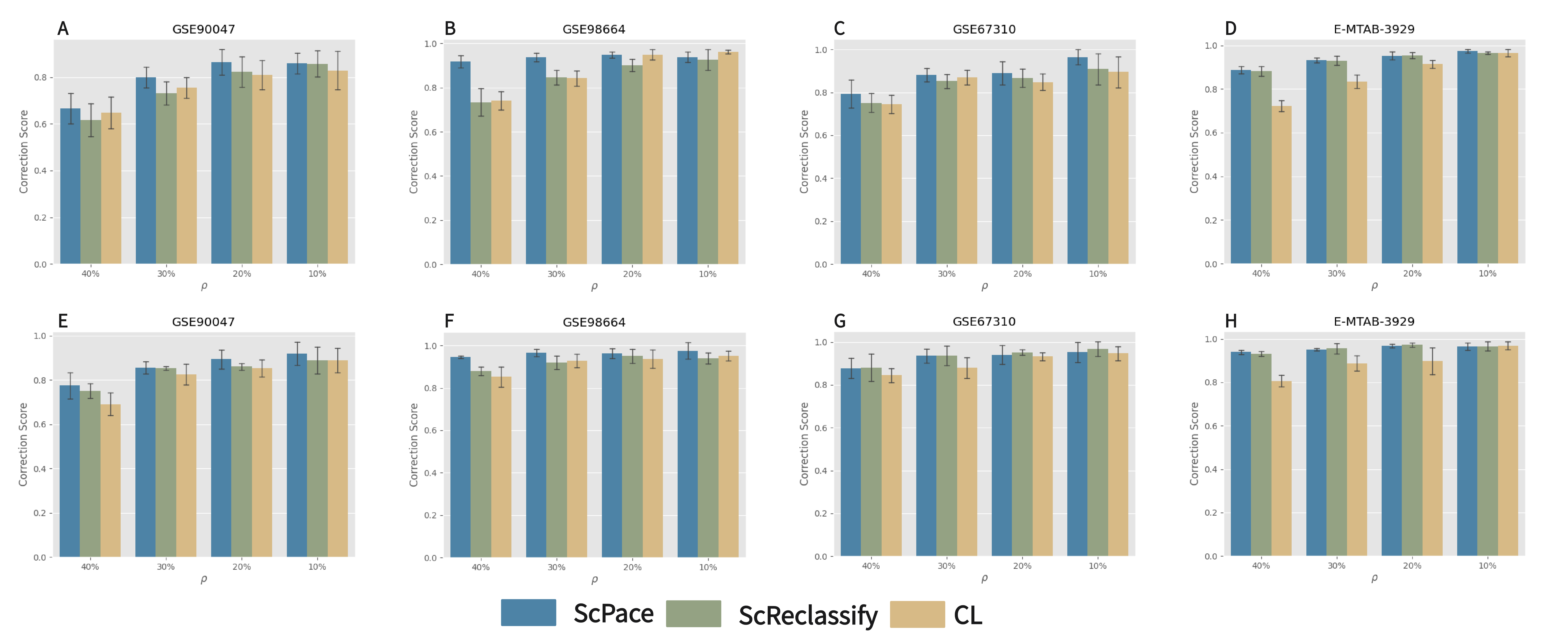}
\caption{Correction score of cross validation with different mislabeling rate on real-time-series ScRNA-seq data (A-D) correction score results with swap mislabeling (E-H) correction score results with random mislabeling}
\label{cor_score_real}
\end{figure}

In comparison to ScReclassify and CL, ScPace consistently demonstrated superior performance across all mislabeling rates and both types of mislabeling. Notably, as the mislabeling rate increased, ScPace maintained strong performance levels, whereas the performance of ScReclassify and CL exhibited a decline. The correction score of the three calibration methods shown in Fig.(\ref{cor_score_sim}) for simulated datasets and Fig.(\ref{cor_score_real}) for real time-series ScRNA-seq demonstrate the superior correction performance of ScPace on most occasions. All the results highlights ScPace's resilience and reliability, particularly in challenging conditions where mislabeling is prevalent. Overall, the results affirm ScPace's effectiveness as a robust classifier across various situations.

Interestingly, we found out that all the classifiers gradually perform better when mislabeling rate decreases. Nevertheless, ScPace's performance is not influenced by the mislabeling rate and it all achieved higher than 90\% score accuracy.

In conclusion, the experimental results show that when assigning time labels to new cells on noisy simulated and real time-series ScRNA-seq datasets, ScPace is robust on both balanced and imbalanced situations, and the experimental results on real time-series ScRNA-seq datasets suggest that ScPace could also be applied to real world situations. This suggest that ScPace could be applied to various noisy time-series ScRNA-seq datasets for timestamps calibration. Moreover, both SVM and Random Forest also performed well when mislabeling rate are under 20\%, but the performance worsens when mislabeling rate increases. which is reasonable that SVM ignores label noise but still sensitive to label noises, and previous study has proved that random forest is also robust to label noises\cite{rf_robust}. Since the base classifier of ScPace is SVM, the experimental results show that compared with traditional SVM, ScPace is more capable of dealing with noisy time-series ScRNA-seq on timestamp automatic annotation.
\subsubsection{Sensitive Analysis}
Improper tuning of hyperparameters can lead to erroneous results in ScPace. To evaluate the impact of hyperparameter variation on model performance, we conducted a sensitivity analysis by systematically adjusting key parameters. These included $C$, the regularization term; $p$, the C-Growing parameter; iter, the number of iterations for running ScPace; $\lambda$, the threshold for updating latent variables. During the analysis of each parameter, the selected hyperparameter was varied while holding others constant. Additionally, the dimensionality reduction method was fixed to PCA throughout the sensitivity analysis.

\begin{figure}[h]   
\centering         
\includegraphics[scale=0.11]{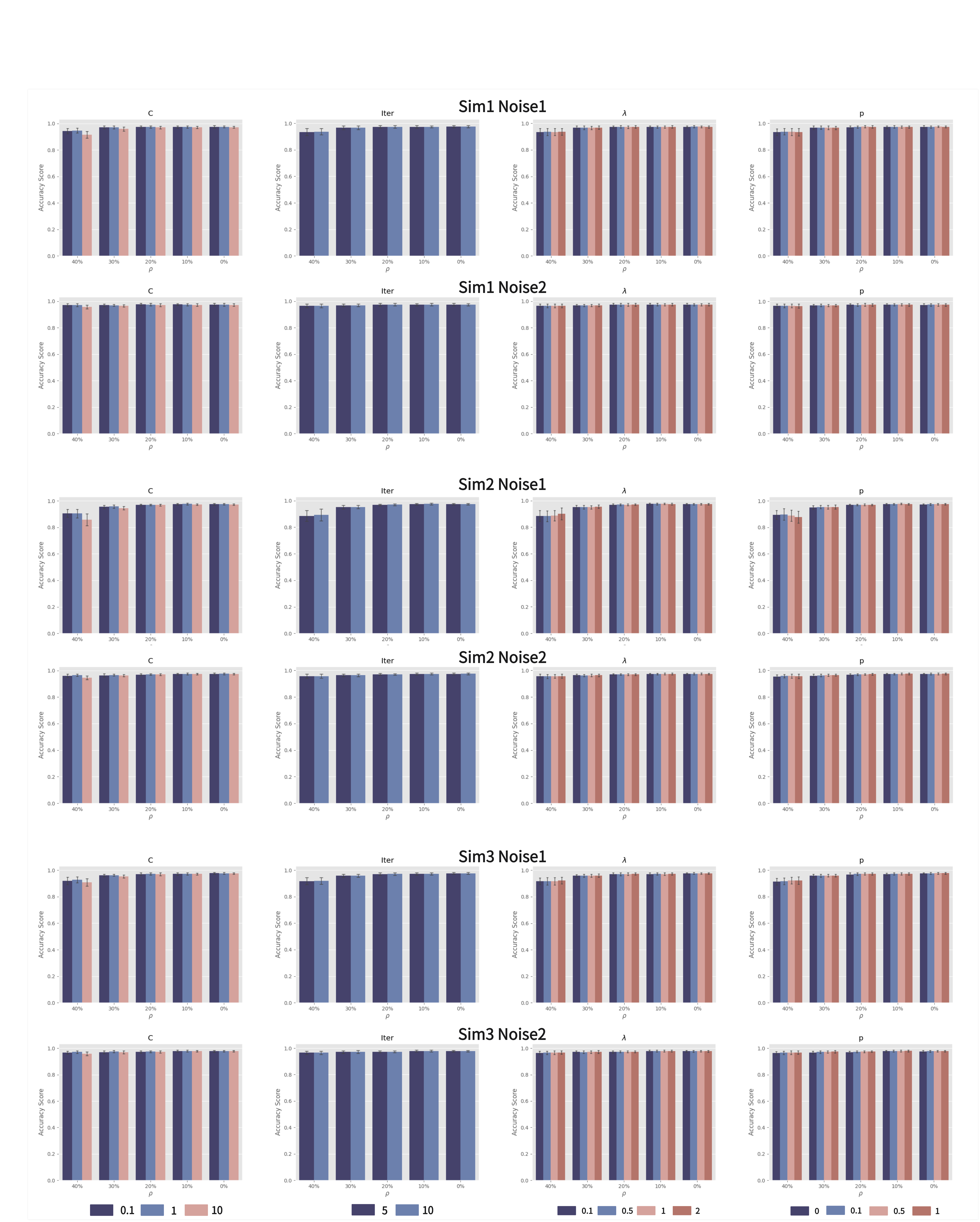}
\caption{Sensitive analysis of four hyperparameters(C, $\lambda$, iter, p) on simulated datasets. Each datasets contains two rows, first row represents the cross validation sensitive analysis on swap mislabeling, second row represents the cross validation sensitive analysis on random mislabeling}
\label{sensitive_simulate}
\end{figure}

\begin{figure}[h]   
\centering         
\includegraphics[scale=0.1]{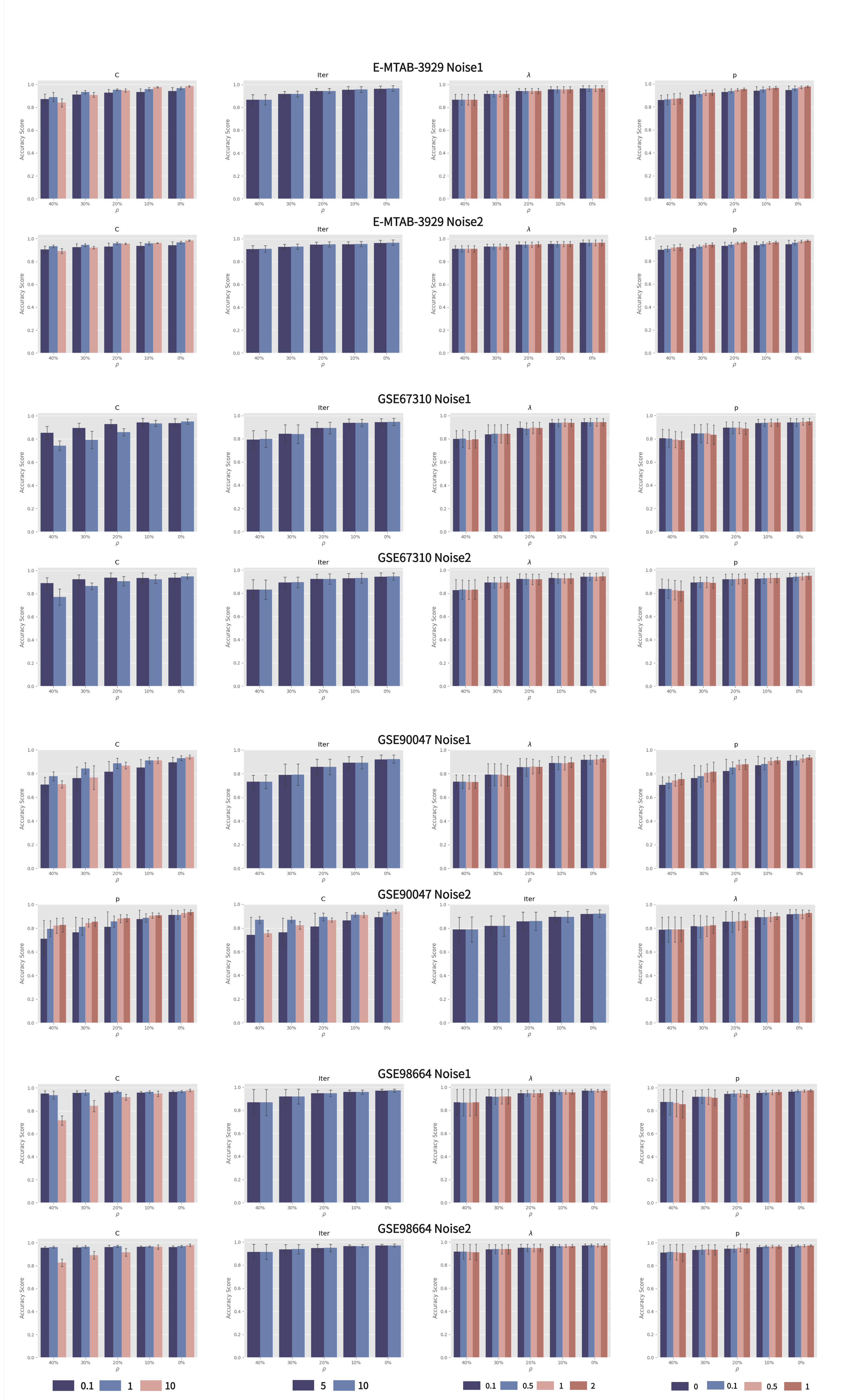}
\caption{Sensitive analysis of four hyperparameters(C, $\lambda$, iter, p) on real time-series ScRNA-seq data. Each datasets contains two rows, first row represents the cross validation sensitive analysis on swap mislabeling, second row represents the cross validation sensitive analysis on random mislabeling}
\label{sensitive_real}
\end{figure}

The results Shown in Fig.(\ref{sensitive_simulate}) for simulated datasets and Fig.(\ref{sensitive_real}) for real time-series ScRNA-seq datasets that besides hyperparameter $C$, the performance of those hyperparameter, including $\lambda$,$Iter$,and $p$, doesn't show significant varying performance. In contrast, we observed that there is a significant varying performance on hyperparameter C on most dataset.

Interestingly, We observed that if the parameter $C$ is set too low or too high, ScPace tends to perform adequately. This sensitivity to hyperparameter tuning highlights the need for a balanced approach to parameter selection, ensuring that the calibration process optimally identifies and mitigates mislabeled cells without unnecessarily compromising the dataset's integrity.

In previous work, ScReclassify utilized dimensional reduction techniques to address the issue of polluted features in ScRNA-seq datasets. To determine if dimensional reduction could likewise enhance classification performance in ScPace, we compared the results of cross-validation with and without dimensional reduction. We applied Principal Component Analysis (PCA) and KernelPCA to both the training and testing sets, analyzing their effects on simulated and real time-series ScRNA-seq datasets. The aim was to assess whether these dimensional reduction methods would improve the classification accuracy of ScPace by reducing noise and retaining only the most informative features.

\begin{figure}[h]   
\centering         
\includegraphics[scale=0.065]{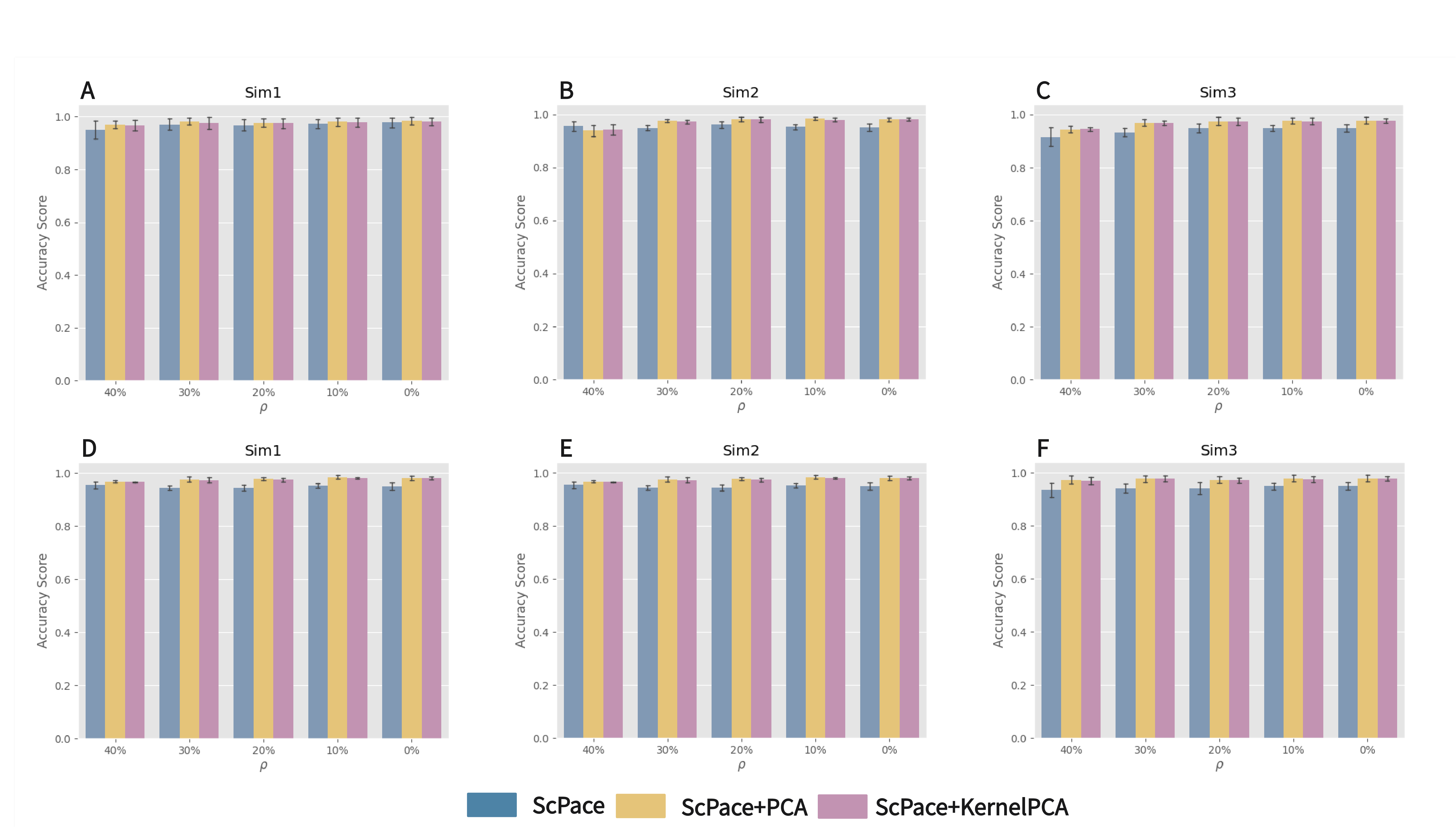}
\caption{Sensitive analysis results using different dimensional reduction technique on simulated datasets (A-C) Cross validation results with different dimensional reduction technique on swap mislabeling (D-F) Cross validation results with different dimensional reduction technique on random mislabeling.}
\label{pca_sim}
\end{figure}

On simulated datasets, our results shown in Fig.(\ref{pca_sim}) indicated that when dimensional reduction was applied to the training datasets, both PCA and KernelPCA outperformed the models without dimensional reduction. This improvement may be attributed to the linear separability of the simulated datasets generated by Splatter, which facilitated better classification outcomes for the reduced datasets.

\begin{figure}[h]   
\centering         
\includegraphics[scale=0.065]{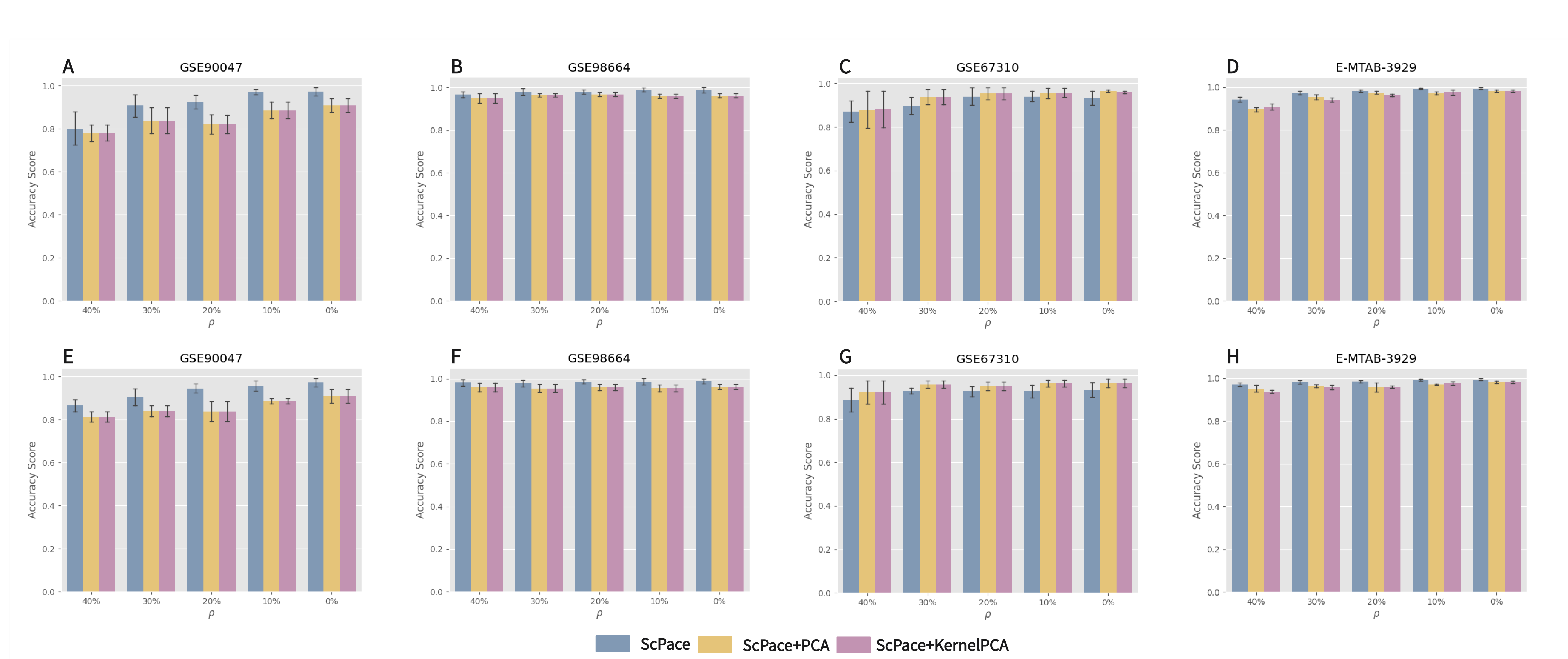}
\caption{Sensitive analysis results using different dimensional reduction technique on real time-series ScRNA-seq datasets (A-D) Cross validation results with different dimensional reduction technique on swap mislabeling(E-H) Cross validation results with different dimensional reduction technique on random mislabeling.}
\label{pca_real}
\end{figure}

However, the performance shown in Fig.(\ref{pca_real}) changed when we examined real time-series ScRNA-seq datasets that are not linearly separable. For instance, in datasets like GSE90047 and GSE98664, models without PCA or KernelPCA dimensional reduction preprocessing tended to outperform those that incorporated these techniques. This suggests that for some complex, non-linearly separable datasets, dimensional reduction may obscure important features rather than enhance classification accuracy.

Conversely, for the dataset GSE63710, we observed that the application of PCA and KernelPCA led to improved performance compared to the non-processed dataset. This indicates that the effectiveness of dimensional reduction can vary significantly depending on the specific characteristics of the dataset, underscoring the need for careful evaluation of preprocessing techniques for each ScRNA-seq analysis scenario.

Additionally, our findings revealed that the experimental results for PCA and KernelPCA were quite similar. This implies that the Kernel PCA methods did not provide a significant boost in classification accuracy compared to PCA.

\subsection{Enhancement of supervised Pseudotime analysis}
\subsubsection{Experimental Setup}
In addition to cross-validation, another reliable way to assess the effectiveness of our proposed method is to evaluate whether the calibrated timestamps can enhance the performance of supervised pseudotime analysis. Previous research has shown that supervised pseudotime analysis is sensitive to the quality of timestamps. To investigate this further, we compared the enhancement results of our method against those achieved using ScReclassify for calibrating timestamps across both simulated and original time-series ScRNA-seq datasets. Note that both ScPace and CL includes two types of timestamp calibration: deletion and reclassification. In contrast, ScReclassify focuses exclusively on reclassification. Therefore, when comparing our method with ScReclassify, we restricted our analysis specifically to the reclassification of noisy simulated and real time-series ScRNA-seq datasets.

We initiated our analysis by applying reclassification through three calibration methods to both the artificially mislabeled simulated datasets and the original real-time series ScRNA-seq datasets (which were not artificially mislabeled).

After completing the reclassification using the three calibration methods on the original noisy simulated and real time-series ScRNA-seq data, we applied Psupertime to the reclassified timestamps from both methods independently and then computed the average training accuracy, testing accuracy and pseudotime value to each cells. To assess the performance of the computed pseudotime, we calculated the Spearman and Kendall correlation between the computed pseudotime value and the calibrated timestamps. Then we applied psupertime to the original timestamps for comparison purpose, seeking whether the calibration of timestamps could enhance the performance.

We also implemented the deletion calibration method of ScPace and CL and compared the results with original timestamps. Through these comparisons, we aimed to evaluate the overall performance of our proposed method in enhancing the results of supervised pseudotime analysis.

\begin{table}
\centering
\setlength{\tabcolsep}{0.1mm}
\begin{tabular}{cccc}
\toprule
Classifiers & Parameter & Description & Values\\
\midrule                                                                                                                                                                  
\textbf{ScPace} & \textbf{C} & The regularization Parameter & \{0.1,1,10\}\\
\textbf{} & \textbf{p} & The C-Growing Parameter & \{0,0.1,0.5,1\}\\
\textbf{} & \textbf{Iter} & Number of iterations to perform ScPace & \{5,10\}\\
\textbf{} & \textbf{$\lambda$} & Threshold for updating the latent variables $v$ & \{0.1,0.5,1,2\}\\

\textbf{ScReclassify} & \textbf{L} & Number of ensembles & \{3,5,10\}\\
\textbf{} & \textbf{$\alpha\%$} & Percentage of samples to select at each iteration & \{0.5,0.8,1\}\\
\textbf{} & \textbf{Iter} & Number of iterations to perform Adasampling & \{3,5,10\}\\
\textbf{} & \textbf{Balance} & Logical flag to if the cell types are balanced & \{0,1\}\\

\textbf{CL} & \textbf{(i,j,k)} & The hidden layer sizes of the network& \{(50,50,50),(50,50,100)\}\\

\textbf{} & \textbf{$\rho(x)$} & The activation function of each neurons & \{$\max{(0,x)}$\}\\

\textbf{Psupertime} & \textbf{sel\_gene} & Method to be used to select interesting genes for Psupertime& \{hvg,all\}\\
\textbf{} & \textbf{error} & Cross-validated accuracy to be used to select model & \{xentropy,class\_error\}\\

\bottomrule
\end{tabular}
\caption{Hyperparameters used for grid search tuning while conducting the calibration of timestamps and supervised pseudotime analysis}
\end{table}

\subsubsection{Experimental Results}

\begin{figure}[h]   
\centering         
\includegraphics[scale=0.095]{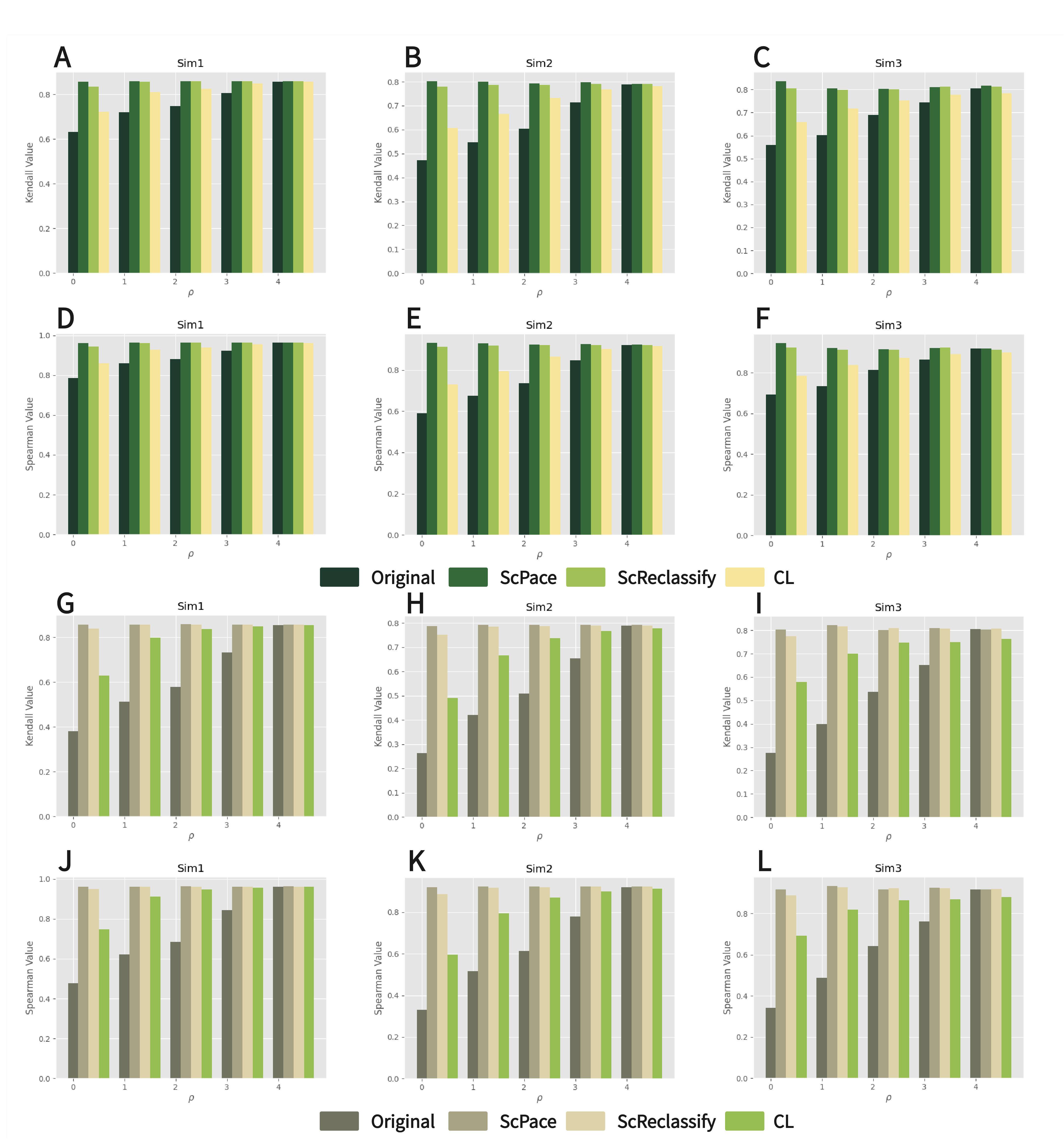}
\caption{Enhancement results of supervised pseudotime analysis on reclassification with two types of mislabeling and two types of metrics (A-C) Enhancement results using Kendall correlation of supervised pseudotime analysis on deletion with swap mislabeling (D-F) Enhancement results using Spearman correlation of supervised pseudotime analysis on deletion with swap mislabeling (G-I) Enhancement results using Kendall correlation of supervised pseudotime analysis with random mislabeling (J-L) Enhancement results using Spearman correlation of supervised pseudotime analysis on deletion with random mislabeling}
\label{reclassify_psupertime}
\end{figure}

The enhancement results of reclassification are illustrated in Fig.(\ref{reclassify_psupertime}), where we observed that ScPace outperformed ScReclassify and CL on the simulated datasets, especially at higher mislabeling rates, such as 40\% and 30\%. However, we noted that Psupertime performed poorly on original mislabeled datasets compared to the calibrated timestamps obtained using ScPace and ScReclassify. This observation suggests that Psupertime lacks robustness when confronted with extremely noisy datasets.

\begin{figure}[h]   
\centering         
\includegraphics[scale=0.1]{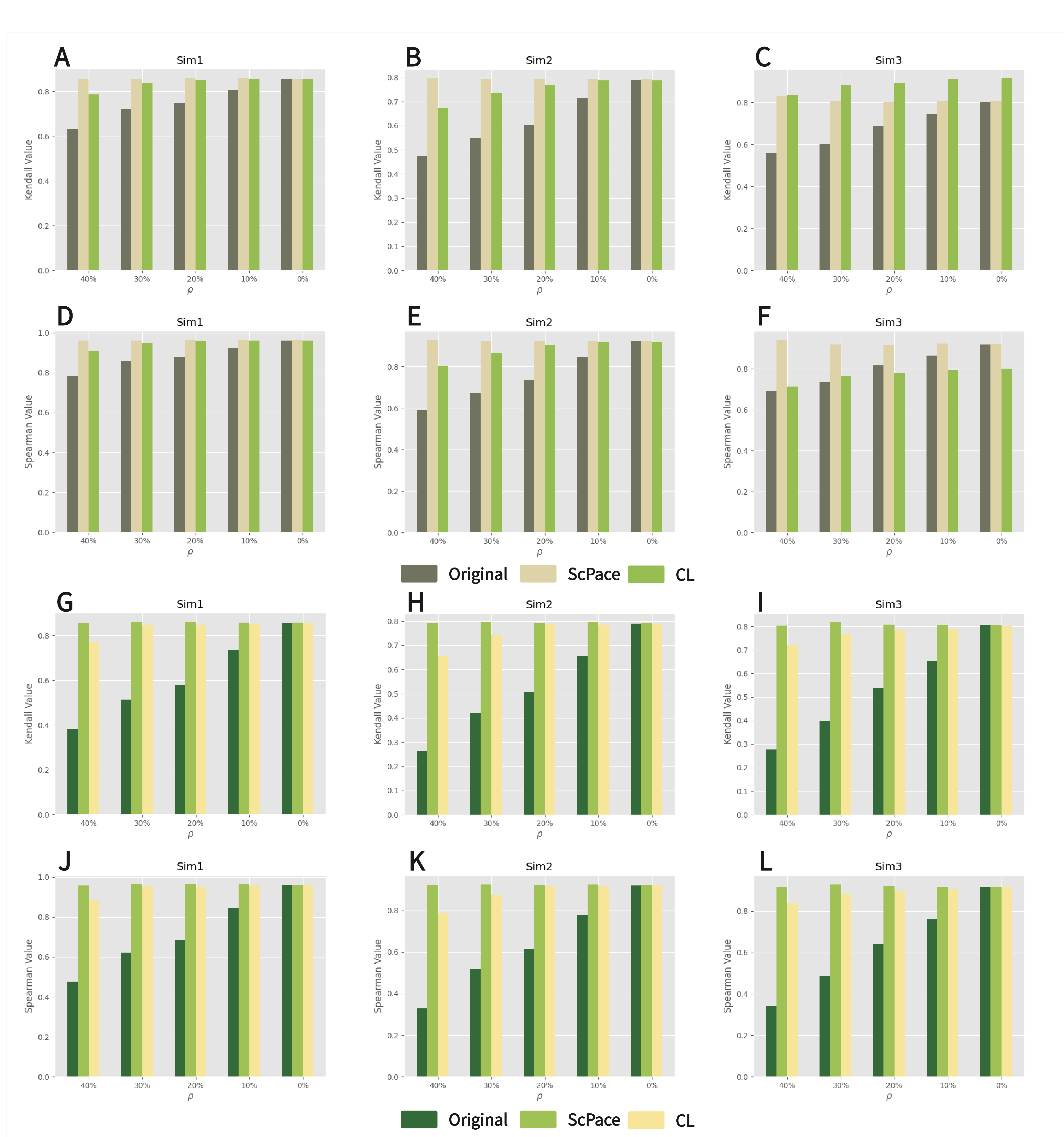}
\caption{Enhancement results of supervised pseudotime analysis on deletion with two types of mislabeling and two types of metrics (A-C) Enhancement results using Kendall correlation of supervised pseudotime analysis with swap mislabeling (D-F) Enhancement results using Spearman correlation of supervised pseudotime analysis with swap mislabeling (G-I) Enhancement results using Kendall correlation of supervised pseudotime analysis with random mislabeling (J-L) Enhancement results using Spearman correlation of supervised pseudotime analysis with random mislabeling}
\label{deletion_psupertime}
\end{figure}

The enhancement results from the deletion method in ScPace, shown in Fig.(\ref{deletion_psupertime}), indicate a clear improvement on the simulated datasets on both types of noise. Moreover, compared with the enhancement results of CL, ScPace performed better. This finding suggests that ScPace effectively removes noisy labeled datasets, leading to enhanced performance in supervised pseudotime analysis. Interestingly, we discovered that Psupertime is sensitive to two types of noise, as the mislabeling rate increases, the correlation values drop significantly.

Moreover, on real time-series ScRNA-seq datasets. ScPace, ScReclassify and CL successfully enhanced the performance of supervised pseudotime analysis through reclassification, however, ScPace demonstrated better performance compared to ScReclassify and CL across all three datasets. This indicates that our proposed method is effective not only in simulated contexts but also in real-world applications, further validating its robustness and utility in supervised pseudotime analysis.

\begin{table}
\centering
\setlength{\tabcolsep}{2.5mm}
\begin{tabular}{cccccc}
\toprule
\textbf{Datasets} &Methods & Mean Training Accuracy & Mean Test Accuracy & Kendall & Spearman\\
\midrule                                                                                                                                                                  
\textbf{GSE90047}&Original & 79\% & 89\% & 0.9096& 0.9822\\
\textbf{}&ScReclassify & 84\% & 91\% & 0.9164& 0.9836\\
 & CL & 83\%& 89\%& 0.9149&0.9851\\
 & ScPace & \textbf{87\%} & \textbf{93\%} & \textbf{0.9225}&\textbf{0.9878}\\

\textbf{GSE67310}&Original & 84\%& 88\%& 0.8659& 0.9640\\
\textbf{}&ScReclassify & 87\% & 92\% & 0.8904& 0.9739\\
 & CL & \textbf{88\%}& 82\%& 0.8662&0.9641\\
 & ScPace & \textbf{88\%} & \textbf{95\%} & \textbf{0.8984}&\textbf{0.9805}\\

\textbf{GSE98664}&Original & 97\%& \textbf{100\%}& 0.8942& 0.9791\\
\textbf{}&ScReclassify & \textbf{98\%} & \textbf{100\%} & 0.8944& 0.9792\\
 & CL & 97\%& 95\%& 0.8945&0.9792\\
 & ScPace & \textbf{98\%} & \textbf{100\%} & \textbf{0.8948}&\textbf{0.9794}\\

\textbf{E-MTAB-3929}&Original & \textbf{97\%}& 94\%& 0.8664& 0.9650\\
\textbf{}&ScReclassify & \textbf{97\%}& 92\%& 0.8672& 0.9654\\
 & CL & \textbf{97\%}& 94\%& 0.8683&0.9662\\
 & ScPace & \textbf{97\%}& \textbf{97\%}& \textbf{0.8688}&\textbf{0.9666}\\

\bottomrule
\end{tabular}
\caption{Results of enhancement of supervised pseudotime analysis using reclassification on original real time-series ScRNA-seq datasets}
\end{table}

We also conducted experiments utilizing the deletion method on real time-series ScRNA-seq. The results showed that the datasets calibrated by ScPace, through the deletion of identified mislabeled cells, significantly enhanced the performance of supervised pseudotime analysis compared with CL. This improvement indicates that the deletion of these cells was not random; rather, it involved a systematic identification of cells that were likely to contribute noise due to mislabeling. Consequently, the removal of these non-representative cells contributed to a clearer and more accurate reconstruction of pseudotime ordering. Thus, our findings suggest that the effective reclassification and deletion of mislabeled cells within the context of time-series ScRNA-seq analyses can lead to substantial improvements in the accuracy and reliability of pseudotime estimations. 
\begin{table}
\centering
\setlength{\tabcolsep}{2.5mm}
\begin{tabular}{cccccc}
\toprule
\textbf{Datasets} &Methods & Mean Training Accuracy & Mean Test Accuracy & Kendall & Spearman\\
\midrule                                                                                                                                      
\textbf{GSE90047}&Original & 79\% & 89\% & 0.9096& 0.9822\\
\textbf{}& CL & 82\%& 77\%& 0.9111& 0.9830\\
\textbf{}&ScPace & \textbf{98\%} & \textbf{100\%} & \textbf{0.9245}& \textbf{0.9886}\\

\textbf{GSE67310}&Original & 84\%& 88\%& 0.8659& 0.9640\\
\textbf{}& CL & 87\%& 90\%& 0.8690& 0.9653\\
\textbf{}&ScPace & \textbf{93\%} & \textbf{97\%} & \textbf{0.8767}& \textbf{0.9695}\\

\textbf{GSE98664}&Original & 97\%& \textbf{100\%}& 0.8942& 0.9791\\
\textbf{}& CL & 96\%& 98\%& 0.8943& 0.9792\\
\textbf{}&ScPace & \textbf{99\%} & \textbf{100\%} & \textbf{0.8945}& \textbf{0.9792}\\

\textbf{E-MTAB-3929}&Original & 97\%& 94\%& 0.8664& 0.9650\\
\textbf{}&CL & \textbf{97\%}& 97\%& 0.8679& 0.9660\\
\textbf{}&ScPace & \textbf{97\%}& \textbf{98\%}& \textbf{0.8700}& \textbf{0.9673}\\
\bottomrule
\end{tabular}
\caption{Best Results comparing the performance of supervised pseudotime analysis between original noisy datasets and deleted datasets conducted by ScPace using real time-series ScRNA-seq datasets}
\end{table}

The experimental findings from both sets of experiments suggest that Psupertime struggles with extreme label noise present in time-series ScRNA-seq datasets. When extreme label noise is prevalent, it hampers the algorithm’s ability to accurately reconstruct pseudotime trajectories, thereby affecting the overall reliability of the analysis. This underlines the importance of employing calibration of timestamps on time-series ScRNA-seq data, such as those offered by ScPace, to calibrate timestamps before conducting supervised pseudotime analyses. 

\section{Case Studies}\label{case}
\subsection{Identification of different maturation state in hepatoblast differentiation} 

We further assessed the detected cells in the GSE90047, which includes two distinct developmental trajectories: one tracing the path from hepatoblast to cholangiocyte and the other from hepatoblast to hepatocyte. Our analysis revealed that the most hepatoblasts present at early timestamps, such as E10.5 and E11.5, were not detected as noisy samples by ScPace. However, we noted some overlap in the cholangiocytes observed at E14.5 and E15.5 along the hepatoblast to cholangiocyte trajectory. Additionally, we found that timestamps between E12.5 and E13.5 exhibited noise across both trajectories, however, label noise was observed to be less pronounced in the hepatoblast to cholangiocyte trajectory compared to the hepatoblast to hepatocyte path. This suggests that, in GSE90047, timestamp noise may arise from varying maturation rates at specific timestamps.

\begin{figure}[h]   
\centering         
\includegraphics[scale=0.065]{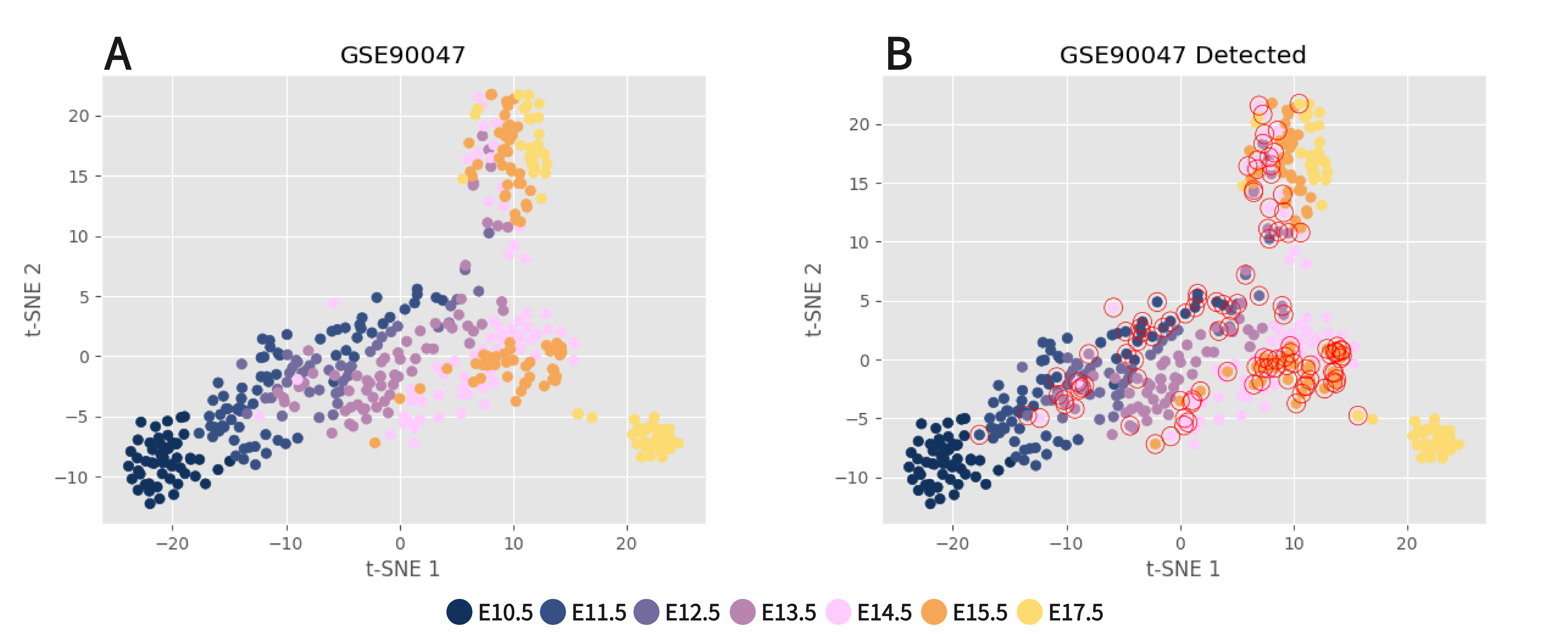}
\caption{(A) Dimensional reduction visualization using TSNE on GSE90047 (B) Potential Mislabeled cells(v=0) detected by ScPace in GSE90047.}
\label{GSE90047_case}
\end{figure}

In Fig. (\ref{GSE90047_case}), the red circles highlight potential mislabeled cells (denoted as $v=0$). Our results show that ScPace successfully detected the majority of overlapping cells in GSE90047, underscoring its capability to identify label noise effectively.
 
Given the challenges posed by these noisy timestamp labels, we do not recommend simply deleting the detected samples in the GSE90047 dataset. Instead, we advocate for reclassification methods to calibrate the cells into more reliable timestamp labels. This approach is preferable because the observed label noises do not appear to stem from technical issues but rather may reflect biological variability. By employing reclassification techniques, we can enhance the integrity of the analysis while retaining valuable information from the dataset. 
\subsection{Illustration of cardiomyocytes developmental status after myocardial infraction}
We performed a primary practice in illustrating cardiomyocytes developmental status after myocardial infraction. The datasets of the cardiomyocytes gathered from GEO database under accession number GSE126128\cite{gse126128}, GSE131181\cite{gse131181} and GSE132658\cite{gse132658}, with 2404 cells and 6 time points(E7.75, E8.25, E9.25, E10.5, E13.5, E16.5) and furthermore preprocessed using Scanpy package. The datasets were then split into training (75\%) and validation (25\%) sets.

\begin{figure}[h]   
\centering         
\includegraphics[scale=0.065]{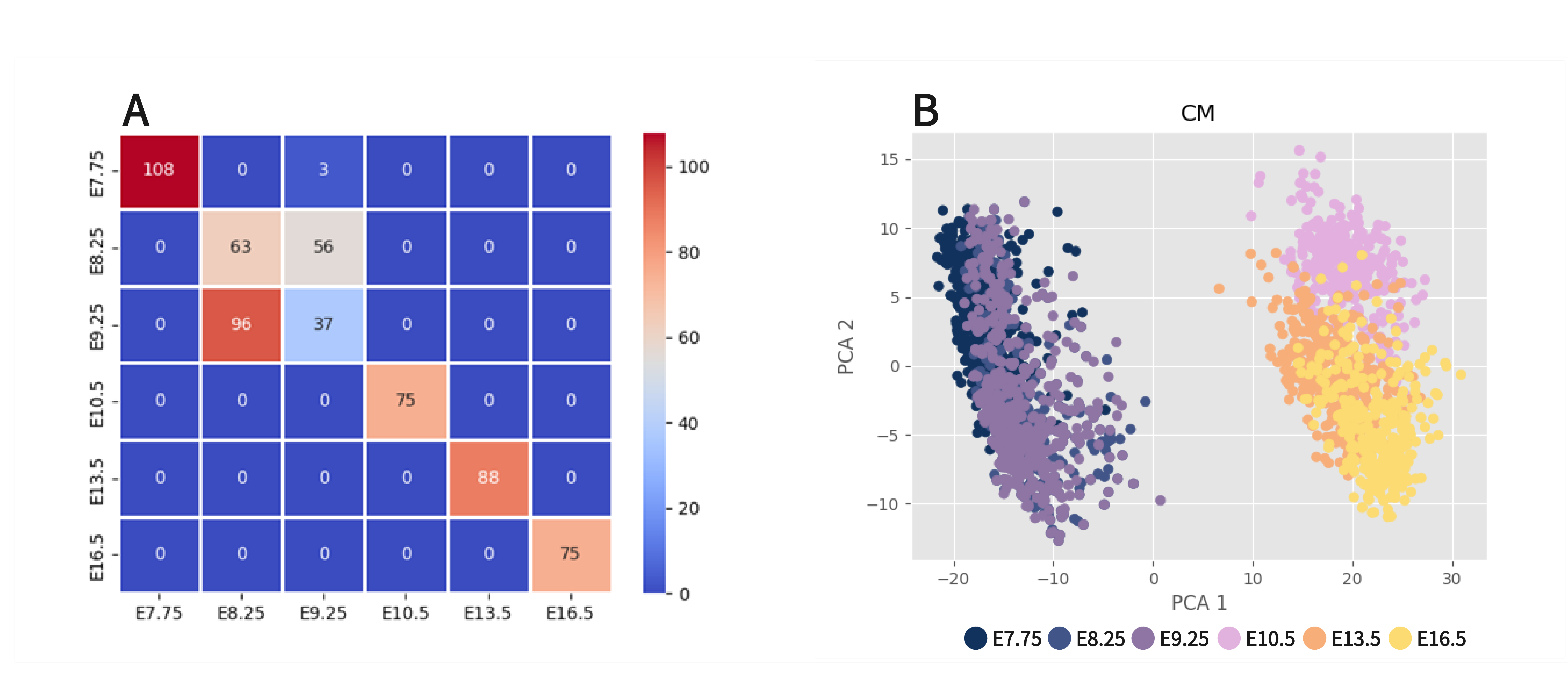}
\caption{(A) Confusion matrix results generated by the validation sets (B) PCA visualization of the cardiomyocyte collected at 6 embryonic days.}
\label{cm_training}
\end{figure}

Based on the training results reflected in the confusion matrix shown in Fig.(\ref{cm_training}), we observed a degree of ambiguity between E8.25 and E9.25. This is likely attributable to the minimal developmental changes occurring during this stage, making it challenging to accurately distinguish between these two time points.

\begin{figure}[h]   
\centering         
\includegraphics[scale=0.055]{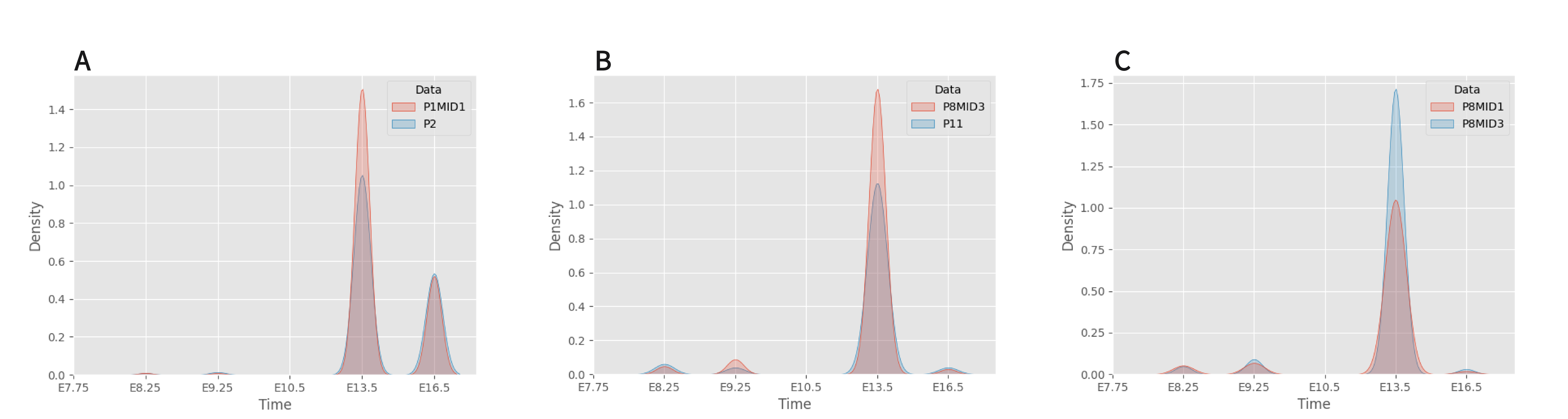}
\caption{Overall procedure of ScPace including data preprocessing, training ScPace and conduct two types of timestamp calibration}
\label{CM_case}
\end{figure}

For testing sets, we gathered 2082 cells from National Center for Biotechnology Information Sequence Read Archive with BioProject accession number PRJNA595398\cite{validation_cm}. In terms of the prediction results from the testing sets shown in Fig.(\ref{CM_case}), we found that the concentration of cardiomyocytes after injury was shifting to the earlier period and the peak at embryonic day 13.5 was specifically increased with recovery process. In addition, our results suggested that obstacles in cardiomyocyte dedifferentiation might occur between status embryonic day 10.5 and day 13.5 due to the density showed little change among earlier periods. Consequently, our results verified cardiomyocyte wound go through the incomplete re-programming process in genomic aspect after myocardial infraction and the number of cardiomyocytes at unmatured status wound increase over time.

\section{Computational timing}\label{timing}
To assess the computational costs of the model on timestamp calibration. The model is trained on a research server with an NVIDIA GPU GeForce RTX 3080 Ti and 12th Gen Intel(R) Core(TM) i7-12700KF CPU with 125Gi . The computational timing of timestamp calibration on both swap mislabeling and random mislabeling are shown in Table.(\ref{noise1_com}) and Table.(\ref{noise2_com}), and the computational timing of real time-series datasets are shown in Table.(\ref{real_timing}) . The results indicate that both ScPace and CL exhibit significantly longer computational times when dimensionality reduction techniques are not employed, compared to ScReclassify, which incorporates PCA in its initial steps. However, we observed that when dimensionality reduction techniques are applied to ScPace, its computational time becomes even shorter than that of ScReclassify.

\section{Discussion}\label{discussion}
In this paper, our study emphasize the importance of assessing the quality of timestamps, which could degrade the performance of timestamp automatic annotation and supervised pseudotime analysis when overlooking the noisy labeled timestamps. Therefore, we proposed ScPace by introducing a latent variable indicator for identifying potential mislabeled cells and calibrate the timestamps of the detected samples. Experiments on both noisy labeled simulated and real time-series ScRNA-seq datasets suggest that both analysis procedures are sensitive to label noises, and the experimental results further demonstrates the effectiveness of ScPace on timestamp automatic annotation and enhancement of supervised pseudotime analysis compared with prior method. 

Nevertheless, while ScPace performs well on large datasets, it does face significant challenges to computational costs. Therefore, future work could focus on developing time consumed optimized methods and algorithms specifically tailored for large and high-dimensional time-series ScRNA-seq datasets, or combining neural network classifier with ScPace, enhancing both efficiency and scalability while maintaining accuracy in calibrating the timestamps.

Another limitation of ScPace arises when the mislabeling rate exceeds 50\%. Under such conditions, where more than half of the cells are mislabeled, the model loses its capacity to reliably identify mislabeled samples. Consequently, future research could prioritize the development of more robust methodologies capable of addressing extreme levels of label noise.
\section{Conclusion}\label{conclusion}
Noisy labeled samples in time-series ScRNA-seq data could degrade the performance of timestamp automatic annotation and supervised pseudotime analysis, further produce unsatisfied biological interpretation. Therefore, in this study, we developed ScPace, a robust timestamp calibration method on time-series ScRNA-seq datasets, which improves the performance of timestamp automatic annotation and further enhance the performance of supervised pseudotime analysis. While ScPace maintain robustness on various type of time-series ScRNA-seq datasets, future work could focus on developing more time consuming algorithm and apply them to evaluate the potential proliferation ability of cardiomyocytes.

\printcredits

\section*{Acknowledgments}

This work was partly supported by the  National Natural Science Foundation of China (62306051,62481540175,82270249), Group Building Scientific Innovation Project for Universities in Chongqing (CXQT21021), Joint Training Base Construction Project for Graduate Students in Chongqing (JDLHPYJD2021016), the Scientific and the Technological Research Program of Chongqing Municipal Education Commission (KJZD-K202400703,KJQN202300718) and the Natural Science Foundation of Chongqing (CSTB2023NSCQ-LZX0092)

\section*{Declaration of competing interest}
The authors declare that they have no known competing financial interests or personal relationships that could have appeared to influence the work reported in this paper. 
\bibliographystyle{elsarticle-num}

\bibliography{Refrences(unmark)}
\newpage

\begin{table}
\centering
\setlength{\tabcolsep}{15mm}
\begin{tabular}{ccc}
\toprule
\textbf{Datasets} &Methods & Training time(Sec)\\
\midrule                                                                                                                                      
\textbf{GSE90047}&CL & 115.07\\
\textbf{}&ScReclassify + SVM & 3.86\\
\textbf{}&ScReclassify + RF & 4.34\\
\textbf{}&ScPace & 71.25\\
\textbf{}&ScPace + PCA & 1.68\\
\textbf{}&ScPace + KernelPCA & 1.12\\

\textbf{GSE67310}&CL & 46.72\\
\textbf{}&ScReclassify + SVM & 2.36\\
\textbf{}&ScReclassify + RF & 1.33\\
\textbf{}&ScPace & 38.64\\
\textbf{}&ScPace + PCA & 0.84\\
\textbf{}&ScPace + KernelPCA & 0.71\\

\textbf{GSE98664}&CL & 102.86\\
\textbf{}&ScReclassify + SVM & 3.91\\
\textbf{}&ScReclassify + RF & 2.35\\
\textbf{}&ScPace & 56.68\\
\textbf{}&ScPace + PCA & 1.46\\
\textbf{}&ScPace + KernelPCA & 1.06\\

\textbf{E-MTAB-3929}&CL & 55.53\\
\textbf{}&ScReclassify + SVM & 6.95\\
\textbf{}&ScReclassify + RF & 6.35\\
\textbf{}&ScPace & 231.27\\
\textbf{}&ScPace + PCA & 2.28\\
\textbf{}&ScPace + KernelPCA & 2.23\\
\bottomrule
\end{tabular}
\caption{Computational time of timestamp calibration on real time-series ScRNA-seq}
\label{real_timing}
\end{table}

\begin{table}
\centering
\setlength{\tabcolsep}{4mm}
\begin{tabular}{ccccccc}
\toprule
\textbf{Datasets} &Methods & 40\%(Sec) & 30\%(Sec) & 20\%(Sec)& 10\%(Sec)& 0\%(Sec)\\
\midrule                                                                                                                                      
\textbf{Sim1}&CL & 50.66& 51.17& 49.55& 45.73& 45.24\\
\textbf{}&ScReclassify + SVM & 6.03& 4.74& 4.92& 5.35& 4.87\\
\textbf{}&ScReclassify + RF & 4.45& 4.49& 4.02& 4.55& 4.70\\
\textbf{}&ScPace & 85.83& 79.76& 76.47& 76.43& 73.88\\
\textbf{}&ScPace + PCA & 2.16& 2.36& 1.86& 1.89& 1.82\\
\textbf{}&ScPace + KernelPCA & 1.79& 1.76& 1.82& 1.51& 1.76\\

\textbf{Sim2}&CL & 79.49& 78.48& 78.87& 87.29& 80.37\\
\textbf{}&ScReclassify + SVM & 6.63& 7.62& 7.67& 8.16& 7.86\\
\textbf{}&ScReclassify + RF & 6.27& 6.07& 6.74& 6.56& 6.88\\
\textbf{}&ScPace & 142.56& 142.74& 154.58& 155.44& 161.57\\
\textbf{}&ScPace + PCA & 2.82& 2.96& 3.02& 3.04& 3.16\\
\textbf{}&ScPace + KernelPCA & 2.66& 2.79& 2.71& 2.85& 3.12\\

\textbf{Sim3}&CL & 151.76& 149.36& 150.43& 138.63& 143.39\\
\textbf{}&ScReclassify + SVM & 14.41& 14.31& 15.38& 14.98& 15.34\\
\textbf{}&ScReclassify + RF& 11.72& 12.07& 12.03& 11.98& 12.12\\
\textbf{}&ScPace & 398.27& 387.96& 370.26& 366.39& 366.09\\
\textbf{}&ScPace + PCA & 7.43& 6.25& 6.17& 6.49& 7.39\\
\textbf{}&ScPace + KernelPCA & 6.06& 6.31& 6.61& 6.14& 6.30\\
\bottomrule
\end{tabular}
\caption{Computational time of timestamp calibration on swap mislabeled simulated time-series ScRNA-seq}
\label{noise1_com}
\end{table}

\begin{table}
\centering
\setlength{\tabcolsep}{4mm}
\begin{tabular}{ccccccc}
\toprule
\textbf{Datasets} &Methods & 40\%(Sec) & 30\%(Sec) & 20\%(Sec)& 10\%(Sec)& 0\%(Sec)\\
\midrule                                                                                                                                      
\textbf{Sim1}&CL & 45.24& 42.48& 47.96& 44.99& 43.63\\
\textbf{}&ScReclassify + SVM & 4.99& 5.24& 5.37& 5.52& 5.58\\
\textbf{}&ScReclassify + RF & 4.12& 4.56& 4.07& 4.36& 4.80\\
\textbf{}&ScPace & 73.87& 72.25& 71.72& 70.29& 70.81\\
\textbf{}&ScPace + PCA & 1.82& 1.73& 1.66& 1.73& 1.78\\
\textbf{}&ScPace + KernelPCA & 1.76& 1.64& 1.58& 1.75& 1.75\\

\textbf{Sim2}&CL & 81.33& 83.82& 80.68& 82.05& 86.57\\
\textbf{}&ScReclassify + SVM & 8.07& 8.58& 8.54& 8.50& 8.32\\
\textbf{}&ScReclassify + RF & 7.44& 6.49& 6.69& 7.08& 6.89\\
\textbf{}&ScPace & 135.55& 129.43& 99.90& 99.88& 100.98\\
\textbf{}&ScPace + PCA & 2.99& 3.21& 2.92& 3.69& 3.29\\
\textbf{}&ScPace + KernelPCA & 2.63& 2.83& 2.97& 3.48& 4.03\\

\textbf{Sim3}&CL & 138.68& 153.44& 155.54& 138.99& 162.81\\
\textbf{}&ScReclassify + SVM & 14.60& 15.22& 15.94& 15.46& 15.87\\
\textbf{}&ScReclassify + RF& 11.89& 11.77& 12.93& 11.68& 11.74\\
\textbf{}&ScPace & 423.12& 403.14& 392.12& 393.69& 393.46\\
\textbf{}&ScPace + PCA & 6.22& 6.54& 6.73& 6.15& 6.56\\
\textbf{}&ScPace + KernelPCA & 6.46& 5.87& 6.14& 5.95& 6.48\\
\bottomrule
\end{tabular}
\caption{Computational time of timestamp calibration on random mislabeled simulated time-series ScRNA-seq}
\label{noise2_com}
\end{table}

\end{document}